\documentclass[10pt]{article}
\usepackage{mathrsfs}
\usepackage{amsfonts}
\usepackage{latexsym,amssymb,amsfonts,amsmath}
\usepackage{graphicx}

\newcommand{\sech}{ {\rm sech} }

\newtheorem{thm}{Theorem}

\newtheorem{lem}[thm]{Lemma}

\newtheorem{defn}[thm]{Definition}
\newtheorem{rem}[thm]{Remark}

\begin{document}
\title{{Cauchy problem and multi-soliton solutions for a two-component short pulse system}}

 \author{Zhaqilao$^{1,3}$\footnote{e-mail:
zhaqilao@imnu.edu.cn},~
  Qiaoyi Hu$^{2,3}$\footnote{e-mail:
huqiaoyi@scau.edu.cn },
~ Zhijun Qiao$^3$\footnote{e-mail: zhijun.qiao@utrgv.edu},
\\$^1$College of Mathematics Science, Inner Mongolia Normal
University,\\ Huhhot, Inner Mongolia 010022, PR China\\
$2$Department of Mathematics, South China
Agricultural University,\\Guangzhou, Guangdong 510642, PR China\\
$^3$School of Mathematical and Statistical Science,\\ University of Texas -- Rio Grande Valley,\\ Edinburg, TX 78539, USA\\
}

\date{}
\maketitle \baselineskip 18pt  {\bf \large{Abstract:}}
In this paper, we study the Cauchy problem and multi-soliton solutions for a two-component short pulse system.
For the Cauchy problem, we first prove the
existence and uniqueness of solution with an estimate of the analytic
lifespan, and then investigate the continuity of the data-to-solution
map in the space of analytic function. For the multi-soliton solutions,
we first derive an $N$-fold Darboux transformation from
the Lax pair of the two-component short
pulse system, which is expressed in terms of the quasideterminant.
Then by virtue of the $N$-fold Darboux transformation we obtain multi-loop and breather soliton solutions.
In particular, one-, two-, three-loop
soliton, and breather soliton solutions are
discussed in details with interesting dynamical interactions and shown through figures.

 \vskip 1mm




\parskip 10pt
\setcounter{equation}{0}
\section{Introduction}

The short pulse equation
\begin{equation}\label{1}
u_{xt}=u+\frac{1}{6}(u^3)_{xx}
\end{equation}
where $u=u(x,t)$ is a real-valued function, representing the magnitude of
the electric field, and the subscripts stand for partial derivatives with respect to $x$ and $t$,
has attracted much attention in the past decades.
This  equation was
derived as a nonlinear model to describe the propagation of
ultra-short pulses in isotropic optical fibers from an approximation to the solutions of the Maxwell's
equations \cite{1}.
It could also be used
to construct integrable differential
equations associated with pseudospherical surfaces \cite{2}.
A numerical
analysis reveals that as the pulse length is shortens, 
equation (\ref{1}) serves as a better approximation to solving the
Maxwell's equation in comparison with the prediction of the
nonlinear Schr\"{o}dinger (NLS) equation \cite{3}.
The short pulse equation (\ref{1})
is  integrable
with the Lax pair \cite{4},
bi-Hamiltonian structure and infinitely many conservation
laws \cite{5}. The Lax pair of the short pulse equation (1) is kind of the WKI-type \cite{Qiao-1992}. Such a kind of Lax pairs, recursion operator, C Neumann and Bargamann constraints to finite dimensional integrable systems, symmetries and transformation to the sine-Gordon equation were  studied in \cite{Qiao-1993,Qiao-1995JMP,6}.  Moreover, 
a suitable hodograph
transformation was found to send the short pulse equation (\ref{1}) to the well-known Sine-Gordon (SG) equation to get multi-loop solitary wave solutions
\cite{9}. 
Therefore, various solutions to the short pulse equation
(\ref{1}) have been obtained, for instance, its loop and
pulse solutions in 
\cite{9},
periodic and solitary wave solutions  
in \cite{10}, two-loop soliton solutions in 
\cite{11}, and bilinear forms, multi-loop solutions, multi-breather and periodic solutions in 
\cite{7,8}.
The loop soliton solutions
to the short pulse equation (\ref{1}) could also be derived from 
a Darboux transformation \cite{12}.

Similar to the case of the NLS equation \cite{15}, there are several different versions to generalize
the short pulse equation (\ref{1}), for instance,  the higher-order nonlinearity corrections in
\cite{16}, and the vector short pulse equations in \cite{17,18}.
Recently, Matsuno proposed a novel multi-component short pulse model \cite{20}, in particular, the following two-component short pulse (2SP) system
\begin{equation}\label{2}
u_{xt}=u+ \frac{1}{2}(uvu_x)_{x},\,\,\,
v_{xt}=v+\frac{1}{2}(uvv_x)_{x}
\end{equation}
was addressed from its multi-component model. Actually, the 2SP system (\ref{2}) was able to be produced from the negative order Wadati-Konno-Ichikawa (WKI) hierarchy in \cite{Qiao-book,QCW-2003}, where the Lax pair for the whole WKI hierarchy and algebaric structure with $r$-matrix were discussed. The regular Wadati-Konno-Ichikawa (WKI) hierarchy and its Lax representations were discussed in \cite{21} and \cite{Qiao-1992}, respectively.
Apparently, 
putting $u=v$ sends (\ref{2}) to the short pulse equation
(\ref{1}). The 2SP system (\ref{2}) is integrable with the following Lax pair \cite{Qiao-book,QCW-2003,20}
\begin{equation}\label{3}
\Psi_x=P\Psi,\,\,\,\,\Psi_t=Q\Psi,
\end{equation}
where
\begin{equation}\label{4}
P=\lambda\left(\begin{array}{cc}  1     & u_x\\[1mm]
                               v_x  & -1
\end{array}
\right),\,\,\,
Q=\dfrac{1}{2}\left(\begin{array}{cc}  0    & -u\\[1mm]
                                       v    &  0
\end{array}
\right)+\dfrac{1}{4\lambda}\left(\begin{array}{cc}  1  & 0\\[1mm]
                                                    0  & -1
\end{array}
\right)+\dfrac{\lambda}{2}\left(\begin{array}{cc}  uv     & uvu_x\\[1mm]
                                                   uvv_x  & -uv
\end{array}
\right).
\end{equation}
Let 
\begin{equation}\label{5}
u=q+ {\rm i} r,\,\,\, v=q- {\rm i} r,
\end{equation}
then the 2SP system (\ref{2}) is cast 
into the following integrable system 
\begin{equation}\label{6}
q_{xt}=q+ \dfrac{1}{2}\left[(q^2+r^2)q_x\right]_x,\,\,\,
r_{xt}=r+
\dfrac{1}{2}\left[(q^2+r^2)r_x\right]_x.
\end{equation}
One- and two-soliton solutions and  breather solutions of equations
(\ref{2}) and (\ref{6}) have been given in terms of pfaffians
by virtue of Hirota's bilinear method in \cite{20}. If $r=0$ $($or
$q=0)$, then the system (\ref{6}) is reduced to the the short pulse
equation (\ref{1}).

In general, once a Lax pair is given one may use it to derive many integrable properties of a nonlinear wave equation such as conservation laws, bi-Hamiltonian structure, Darboux transformation \cite{22,23,24,25,26,27} etc.
On the other hand, the study of analyticity for nonlinear wave equations is another important field in the theory  of partial differential equations. 
For instance, the hydrodynamics of Euler equations was initiated by Ovsyannikov \cite{O1,O2} and later developed with a further study in \cite{CE2,Nr,Ns,Tre1} and in \cite{B-G1,B-G2} where the approach is based on a contraction type argument in a suitable scale of Banach spaces.
The analyticity of the Cauchy problem for the two-component Camassa-Holm shallow water and the two-component Hunter-Saxton systems were studied in \cite{YY1,YY2}.
Recently, Barostichi, Himonas and Petronilho \cite{B-H-P1} established the well-posedness for a class of nonlocal evolution equations 
in spaces of analytic functions. Furthermore, they proved a Cauchy-Kovalevsky theorem for a generalized Camassa-Holm equation ($g-kbCH$) in \cite{B-H-P}. Very Recently, Luo and Yin investigated the Gevrey regularity and analyticity for a class of Camassa-Holm type systems \cite{Luo}.
Thus, an amazing topic is to study the analytic solutions for the 2SP system (\ref{2}).
It is worthy to point out that our approach is strongly motivated from the Cauchy-Kovalevsky type results in \cite{B-G2,B-H-P1}.


The present paper is two fold: studying Darboux
transformation and Cauchy problem of the 2SP system (\ref{2}).
The whole paper is organized as follows. 
In
section 2, we adopt a hodograph transformation to transform the 2SP equation
(\ref{2}) to another nonlinear partial differential equation,
which is a Lax integrable system belonging to the Heisenberg
ferromagnet (HF) hierarchy \cite{24,28}. In section 3, we define a Darboux transformation in terms of Darboux matrix operator, and then
provide a detailed proof for the Darboux transformation and its quasideterminant
representation of the $N$-fold case. In section 4, based on scalar solutions of the
Lax pair, we
construct a $N$-fold Darboux transformation to derive the explicit  multi-loop soliton and breather soliton solutions with their dynamical interactions. Three special examples are discussed details and shown through their graphs.
Starting from section 5, we study the Cauchy problem for the 2SP system (\ref{2}). To do so,
some preliminary results are first provided including the abstract Ovsyannikov type theorem and the basic properties of the analytic space $G^{\delta,s}$. In section 5.1, we prove the existence and
uniqueness of analytic solution with an estimate about the analytic lifespan, and 
in section 5.2, we investigate the continuity of the data-to-solution
map in spaces of analytic functions. Last section concludes the paper with a
brief summary and a future outlook.


\section{Hodograph transformation}

In this section, we 
show how the Lax pair (\ref{3}) is related by a hodograph transformation to a negative order flow
in the HF hierarchy. As per 
\cite{20}, let us introduce
a dependent variable $w$ satisfying
\begin{equation}\label{104}
w^2=1+v_x\,u_x,
\end{equation}
which is able to send the 2SP system (\ref{2}) to
the following form of conservation law
\begin{equation}\label{105}
w_t=(\frac{1}{2}uvw)_x.
\end{equation}
We then define a hodograph transformation
$HT: \ (x,t)\rightarrow(y,\tau)$ by means of
\begin{equation}\label{106}
dy=wdx+\frac{1}{2}uvwdt,\,\,\,d\tau=dt,
\end{equation}
or equivalently
\begin{equation}\label{7}
\frac{\partial}{\partial x}=w\frac{\partial}{\partial
y},\,\,\,\frac{\partial}{\partial t}=\frac{\partial}{\partial
\tau}+\frac{1}{2}uvw\frac{\partial}{\partial y}.
\end{equation}

The 2SP system (\ref{2}) can now be
expressed in terms of new variables in the following form
\begin{equation}\label{8}
x_{y\tau}=-\frac{1}{2}(uv)_y,\,\,\,u_{y\tau}=x_y
u,\,\,\,v_{y\tau}=x_y v.
\end{equation}
System (\ref{8}) arises as a zero curvature equation
$U_\tau-V_{y}+[U,V]=0$, which is exactly the compatibility condition for
the following Lax pair
\begin{equation}\label{10}
\Psi_y=U(y,\tau;\lambda)\Psi,\,\,\,\Psi_\tau=V(y,\tau;\lambda)\Psi,
\end{equation}
where
\begin{equation}\label{11}
U(y,\tau;\lambda)=\lambda\partial_yR,\,\,\,V(y,\tau;\lambda)=S+\frac{1}{\lambda}W,
\end{equation}
and
\begin{equation}\label{12}
R=\left(\begin{array}{cc}  x  & u\\[1mm]
                                  v  & -x
\end{array}
\right),\,\,\,
S=\frac{1}{2}\left(\begin{array}{cc}  0   & -u\\[1mm]
                                      v   &  0
\end{array}
\right),\,\,\,
W=\frac{1}{4}\left(\begin{array}{cc}  1    &   0\\[1mm]
                                      0    &  -1
\end{array}
\right).
\end{equation}
Thus, system (\ref{8}) is actually a negative order flow in the HF hierarchy.

\section{Darboux transformation}
Based on the Lax pair (\ref{10}) of the integrable equation
(\ref{8}), let us consider the following Darboux transformation:
\begin{equation}\label{13}
\Psi[1]=D\Psi,
\end{equation}
where $D$ is a Darboux matrix and $\Psi[1]$ recovers the form of the
Lax pair (\ref{10})
\begin{equation}\label{14}
\Psi[1]_y=U[1]\Psi[1]=\lambda(\partial_yR[1])\Psi[1],\,\,\,U[1]=(D_y+DU)D^{-1},
\end{equation}
\begin{equation}\label{15}
\Psi[1]_\tau=V[1]\Psi[1]=(S[1]+\frac{1}{\lambda}W[1])\Psi[1],\,\,\,V[1]=(D_\tau+DV)D^{-1},
\end{equation}
with
\begin{equation}\label{16}
R[1]=\left(\begin{array}{cc}  x[1]  & u[1]\\[1mm]
                                     v[1]  & -x[1]
\end{array}
\right),\,\,\,
S[1]=\frac{1}{2}\left(\begin{array}{cc}  0     & -u[1]\\[1mm]
                                         v[1]  &  0
\end{array}
\right),\,\,\,
W[1]=\frac{1}{4}\left(\begin{array}{cc}  1    &   0\\[1mm]
                                         0    &  -1
\end{array}
\right),
\end{equation}
and $x[1]$, $u[1]$ and $v[1]$ being new solutions to the equation
(\ref{8}), that is,
\begin{equation}\label{17}
x[1]_{y\tau}=-\frac{1}{2}(u[1]v[1])_y,\,\,\,u[1]_{y\tau}=x[1]_yu[1],\,\,\,v[1]_{y\tau}=x[1]_yv[1].
\end{equation}
In order to make the covariance of the Lax pair (\ref{10}) under the
Darboux transformation (\ref{13}), the key step is to find an
appropriate Darboux matrix $D$ such that $R[1]$, $S[1]$  and $W[1]$
in Eq. (\ref{16}) have the same form as $R$, $S$  and $W$ in
Eq. (\ref{12}). Meanwhile, the old potentials $x$, $u$ and
$v$ in $R$, $S$ are mapped into the new potential $x[1]$, $u[1]$ and
$v[1]$ in $R[1]$, $S[1]$.

For the Lax pair(\ref{10}), let us define the following Darboux matrix
\begin{equation}\label{19}
\Psi[1]= D\Psi\equiv(\lambda^{-1}I-M)\Psi,
\end{equation}
where $I$ is the $2\times2$ identity matrix and
\begin{equation}\label{20}
M=H\Lambda^{-1}H^{-1},
\end{equation}
with
\begin{equation}\label{21}
\Lambda^{-1}=\left(\begin{array}{cc}  \frac{1}{\lambda_1}     & 0\\[1mm]
                                      0 &  \frac{1}{\lambda_2}
\end{array}
\right),\,\,\,H=(\Psi(\lambda_1)|e_1\rangle,\,\Psi(\lambda_2)|e_2\rangle).
\end{equation}
In equation (\ref{21}), $|e_1\rangle$ and $|e_2\rangle$ are two
constant vectors, and $\Psi(\lambda_1)$ and $\Psi(\lambda_2)$
are two fundamental-matrix solutions 
of the Lax pair (\ref{10}) corresponding to the eigenvalues $\lambda_1$
and $\lambda_2$, respectively. Thus, the Lax pair (\ref{10}) can be
rewritten in the following matrix form 
\begin{equation}\label{22}
H_y=R_yH\Lambda,
\end{equation}
\begin{equation}\label{23}
H_\tau=SH+WH\Lambda^{-1},
\end{equation}
where $H$ is the matrix solution of the Lax form (\ref{10})
corresponding to the eigen-matrix $\Lambda$ consisting of distinct eigenvalues.

Using the above facts, we may have the following
propositions.

{\bf Proposition 1.} Under the Darboux transformation (\ref{19}), the matrix $R[1]$ given by the first equation
of (\ref{16}) has the same form as  $R$ in the first equation of
(\ref{12})  with 
the following condition
\begin{equation}\label{24}
R[1]=R-M,
\end{equation}
where the matrix $M$ satisfies
\begin{equation}\label{25}
M_yM=[R_y,M].
\end{equation}

{\bf Proof 1.} Let $R-R[1]=M$. 
Let us show that under the Darboux transformation (\ref{19})$M=H\Lambda^{-1}H^{-1}$
is a solution of equation (\ref{25}). Taking derivative with respect to
$y$ on both sides of $M=H\Lambda^{-1}H^{-1}$ yields: 
\begin{equation}\label{26}
\begin{array}{ll}
M_y=H_y\Lambda^{-1}H^{-1}+H\Lambda^{-1}H^{-1}_y\\[2mm]
\mbox{}\hskip 0.5cm=R_y-H\Lambda^{-1}H^{-1}H_yH^{-1}\\[2mm]
\mbox{}\hskip 0.5cm=R_y-H\Lambda^{-1}H^{-1}R_yH\Lambda H^{-1}\\[2mm]
\mbox{}\hskip 0.5cm=R_y-MR_yM^{-1},
\end{array}
\end{equation}
which is equivalent to the condition (\ref{25}). The proof is thus completed.

{\bf Proposition 2.} Under the
Darboux transformation (\ref{19}), the matrices $S[1]$ and $W[1]$ given by the
second and third equations of (\ref{16}) have the same form as $S$
and $W$ in the second and third equations of (\ref{12}) with
the following conditions
\begin{equation}\label{27}
S[1]=S+[W,M], \
W[1]=W,
\end{equation}
where the matrix $M$ satisfies
\begin{equation}\label{29}
M_\tau=[S,M]+[W,M]M.
\end{equation}

{\bf Proof 2.} Let $S[1]-S=[W,M]$, where $M$ is to be determined.
Apparently, the third equation of (\ref{16}) and
(\ref{12}) imply $W[1]=W$. 
Let us now prove that under the Darboux transformation (\ref{19})$M=H\Lambda^{-1}H^{-1}$
is a solution of equation (\ref{29}).
Taking derivative with respect to
$\tau$ on both sides of $M=H\Lambda^{-1}H^{-1}$ leads to: 
\begin{equation}\label{30}
\begin{array}{ll}
M_\tau=H_\tau\Lambda^{-1}H^{-1}+H\Lambda^{-1}H^{-1}_\tau\\[2mm]
\mbox{}\hskip 0.5cm=SH\Lambda^{-1}H^{-1}+WH\Lambda^{-2}H^{-1}-H\Lambda^{-1}H^{-1}S-H\Lambda^{-1}H^{-1}WH\Lambda^{-1}H^{-1}\\[2mm]
\mbox{}\hskip 0.5cm=SM+WM^2-MS-MWM\\[2mm]
\mbox{}\hskip 0.5cm=[S,M]+[W,M]M,
\end{array}
\end{equation}
which is exactly Eq. (\ref{29}). This completes the proof of the
proposition 2.

The interesting thing is to iterate the above
Darboux transformation (\ref{19}) $N$-times to generate a 
quasideterminant representation of the so-called $N$-fold Darboux
transformation. To do so, we will adopt the notion of
the following quasideterminant about the $n\times n$ matrix $D$ introduced by Gelfand and Retakh \cite{29}:
\begin{equation}\label{31}
\begin{array}{ll}
\left|\begin{array}{cc}  A    &   B\\[1mm]
                         C    &   \boxed{D}
\end{array}
\right|=D-CA^{-1}B,
\end{array}
\end{equation}
where $A$, $B$ and $C$ are $n\times n$ matrices and $A$ is
invertible. From Eqs. (\ref{19}), (\ref{24}) and (\ref{27}), the one-fold Darboux transformation (\ref{19}) and the
matrices $R$, $S$ and $W$ can be expressed in terms of
quasideterminants as follows
\begin{eqnarray}\label{32}
\Psi[1]&=&(\lambda^{-1}I-H\Lambda^{-1}H^{-1})\Psi
=\left|
\begin{array}{cc}  H               &   \Psi\\[1mm]
                   H\Lambda^{-1}   &   \boxed{\lambda^{-1}\Psi}
\end{array}
\right|,\\
\label{33}
R[1]&=&R-H\Lambda^{-1}H^{-1} =R+\left|
\begin{array}{cc}  H               &   I\\[1mm]
                   H\Lambda^{-1}   &   \boxed{0}
\end{array}
\right|,\\
\label{34}
S[1]&=&S-[W,-H\Lambda^{-1}H^{-1}]=S-\left[W,\left|
\begin{array}{cc}  H               &   I\\[1mm]
                   H\Lambda^{-1}   &   \boxed{0}
\end{array}
\right|\right],\\
\label{35}
W[1]=W,
\end{eqnarray}
where $0$ is an null matrix. The $N$ times iteration of the Darboux
transformation  gives the quasidetermant matrix solution to
the 2SP system (\ref{2}). Let $H_k$
$(k=1,2,\ldots,N)$ be the invertible matrix solution to the
Lax system (\ref{10}) corresponding to the eigenvalue $\Lambda=\Lambda_k$
$(k=1,2,\ldots,N)$. Then the $N$-fold Darboux transformation 
for the Lax pair (\ref{10}) can be written in the form of 
\begin{equation}\label{36}
\Psi[N]=\left|
\begin{array}{cccccccc}  H_1                &H_2               &\cdots &H_N               &\Psi             \\[1mm]
                         H_1\Lambda_1^{-1}  &H_2\Lambda_2^{-1} &\cdots &H_N\Lambda_N^{-1} & \lambda^{-1}\Psi\\[1mm]
                         H_1\Lambda_1^{-2}  &H_2\Lambda_2^{-2} &\cdots &H_N\Lambda_N^{-2} & \lambda^{-2}\Psi\\[1mm]
                         \vdots             &\vdots            &\ddots &\vdots            & \vdots          \\[1mm]
                         H_1\Lambda_1^{-N}  &H_2\Lambda_2^{-N} &\cdots &H_N\Lambda_N^{-N} &\boxed{\lambda^{-N}\Psi}
\end{array}
\right|.
\end{equation}
The $N$-fold Darboux transformation sends the matrices $R$, $S$
and $W$ to the following forms
\begin{equation}\label{37}
R[N]=R+\left|
\begin{array}{cccccccc}  H_1                    &H_2                   &\cdots &H_N                   &0      \\[1mm]
                         H_1\Lambda_1^{-1}      &H_2\Lambda_2^{-1}     &\cdots &H_N\Lambda_N^{-1}     &0      \\[1mm]
                         H_1\Lambda_1^{-2}      &H_2\Lambda_2^{-2}     &\cdots &H_N\Lambda_N^{-2}     &0      \\[1mm]
                         \vdots                 &\vdots                &\ddots &\vdots                &\vdots \\[1mm]
                         H_1\Lambda_1^{-(N-1)}  &H_2\Lambda_2^{-(N-1)} &\cdots &H_N\Lambda_N^{-(N-1)} &I\\[1mm]
                         H_1\Lambda_1^{-N}      &H_2\Lambda_2^{-N}     &\cdots &H_N\Lambda_N^{-N}     &\boxed{0}
\end{array}
\right|,
\end{equation}
\begin{equation}\label{38}
S[N]=S-\left[W,\left|
\begin{array}{cccccccc}  H_1                    &H_2                   &\cdots &H_N                   &0      \\[1mm]
                         H_1\Lambda_1^{-1}      &H_2\Lambda_2^{-1}     &\cdots &H_N\Lambda_N^{-1}     &0      \\[1mm]
                         H_1\Lambda_1^{-2}      &H_2\Lambda_2^{-2}     &\cdots &H_N\Lambda_N^{-2}     &0      \\[1mm]
                         \vdots                 &\vdots                &\ddots &\vdots                &\vdots \\[1mm]
                         H_1\Lambda_1^{-(N-1)}  &H_2\Lambda_2^{-(N-1)} &\cdots &H_N\Lambda_N^{-(N-1)} &I\\[1mm]
                         H_1\Lambda_1^{-N}      &H_2\Lambda_2^{-N}     &\cdots &H_N\Lambda_N^{-N}     &\boxed{0}
\end{array}
\right|\right],
\end{equation}
\begin{equation}\label{39}
W[N]=W.
\end{equation}

\section{Explicit loop soliton and multi-loop soliton solutions}

According to Propositions 1 and 2, we will use the $N$-fold
Darboux transformation to construct scalar solutions of the Lax system
(\ref{10}), and further give multi-loop and breather soliton solutions with their
interactional dynamics for the 2SP system
(\ref{2}).

Let $(\psi_{(1,k)},\,\phi_{(1,k)})^{T}$ and
$(\psi_{(2,k)},\,\phi_{(2,k)})^{T}$  be two linearly independent solutions to the Lax system (\ref{10}) 
associated with the eigenvalue
$\lambda_k$ $(k=1,2,\ldots,N)$, then its general solution 
has the following form
\begin{equation}\label{40}
\left(\begin{array}{cc}  \psi_k  \\[1mm]
                         \phi_k
\end{array}
\right)=\left(\begin{array}{cc}  \psi_{(1,k)}   &   \psi_{(2,k)}\\[1mm]
                                 \phi_{(1,k)}   &   \phi_{(2,k)}
\end{array}
\right)|e_k\rangle,
\end{equation}
where $|e_k\rangle$ is a two-dimensional constant column vector. Without loss of
generality, we choose $|e_k\rangle=(\mu_k,\,1)^{T}$, where
$\mu_k\in\mathbb{R}$\,\,$(k=1,2,\ldots,N)$.

In order to construct the $N$-fold Darboux transformation of the Lax
system ({\ref{10}) in the form of
\begin{equation}\label{411}
\Psi[N]=D\Psi=\left(\begin{array}{cc}  A   &   B\\[1mm]
                                       C   &   E
                                       \end{array}
\right)\Psi,
\end{equation}
let us define entries of the Darboux matrix $D$ as follows
\begin{equation}\label{412}
A=\lambda^{-N}+\sum^{N-1}_{k=0}A_k\lambda^{-k},\,\,B=\sum^{N-1}_{k=0}B_k\lambda^{-k},\,\,
C=\sum^{N-1}_{k=0}C_k\lambda^{-k},\,\,E=\lambda^{-N}+\sum^{N-1}_{k=0}E_k\lambda^{-k},
\end{equation}
where $A_k$, $B_k$, $C_k$, and $E_{k}$ $(k=0,1,2,\ldots,N-1)$ are
functions of $y$ and $\tau$. Then, we have the following $N$-fold Darboux transformation on
the scalar solutions to the Lax system (\ref{10}).

{\bf Proposition 1$^{\prime}$.} Under the Darboux transformation (\ref{411}), the matrix $R[1]$ given by the first equation of (\ref{16}) has the same form as  $R$ in the first
equation of (\ref{12}), while the transformation sends the old potentials $x$, $u$,
and $v$ to the following new ones
\begin{equation}\label{413}
x[N]=x[0]+\dfrac{\Delta_{A_{N-1}}}{\Delta_{N-1}},\,\,\,
u[N]=u[0]+\dfrac{\Delta_{B_{N-1}}}{\Delta_{N-1}},\,\,\,
v[N]=v[0]+\dfrac{\Delta_{C_{N-1}}}{\Delta_{N-1}},
\end{equation}
where
\begin{equation}\label{424}
\Delta_{N-1}=\left|
\begin{array}{ccccccccccccccccc}  \psi_1       &\psi_1^{(1)}     &\cdots &\psi_1^{(N-2)}   &\psi_1^{(N-1)}         &\phi_1     &\phi_1^{(1)}    &\cdots &\phi_1^{(N-2)}           &\phi_1^{(N-1)}\\[1mm]
                                  \psi_2       &\psi_2^{(1)}     &\cdots &\psi_2^{(N-2)}   &\psi_2^{(N-1)}         &\phi_2     &\phi_2^{(1)}    &\cdots  &\phi_2^{(N-2)}          &\phi_2^{(N-1)}\\[1mm]
                                  \vdots       &\vdots           &\ddots &\vdots           &\vdots                 &\vdots     &\vdots          &\ddots  &\vdots                  &\vdots\\[1mm]
                                  \psi_{2N}    &\psi_{2N}^{(1)}  &\cdots &\psi_{2N}^{(N-2)}&\psi_{2N}^{(N-1)}      &\phi_{2N}  &\phi_{2N}^{(1)} &\cdots  &\phi_{2N}^{(N-2)}       &\phi_{2N}^{(N-1)}
\end{array}
\right|,
\end{equation}

\begin{equation}\label{425}
\Delta_{A_{N-1}}=\left|
\begin{array}{ccccccccccccccccc}   \psi_1       &\psi_1^{(1)}     &\cdots &\psi_1^{(N-2)}   &-\psi_1^{(N)}         &\phi_1     &\phi_1^{(1)}    &\cdots &\phi_1^{(N-2)}           &\phi_1^{(N-1)}\\[1mm]
                                   \psi_2       &\psi_2^{(1)}     &\cdots &\psi_2^{(N-2)}   &-\psi_2^{(N)}         &\phi_2     &\phi_2^{(1)}    &\cdots  &\phi_2^{(N-2)}          &\phi_2^{(N-1)}\\[1mm]
                                   \vdots       &\vdots           &\ddots &\vdots           &\vdots                 &\vdots     &\vdots          &\ddots  &\vdots                  &\vdots\\[1mm]
                                   \psi_{2N}    &\psi_{2N}^{(1)}  &\cdots &\psi_{2N}^{(N-2)}&-\psi_{2N}^{(N)}      &\phi_{2N}  &\phi_{2N}^{(1)} &\cdots  &\phi_{2N}^{(N-2)}       &\phi_{2N}^{(N-1)}
\end{array}
\right|,
\end{equation}

\begin{equation}\label{426}
\Delta_{B_{N-1}}=\left|
\begin{array}{ccccccccccccccccc}  \psi_1       &\psi_1^{(1)}     &\cdots &\psi_1^{(N-2)}   &\psi_1^{(N-1)}         &\phi_1     &\phi_1^{(1)}    &\cdots &\phi_1^{(N-2)}           &-\psi_1^{(N)}\\[1mm]
                                  \psi_2       &\psi_2^{(1)}     &\cdots &\psi_2^{(N-2)}   &\psi_2^{(N-1)}         &\phi_2     &\phi_2^{(1)}    &\cdots  &\phi_2^{(N-2)}          &-\psi_2^{(N)}\\[1mm]
                                  \vdots       &\vdots           &\ddots &\vdots           &\vdots                 &\vdots     &\vdots          &\ddots  &\vdots                  &\vdots\\[1mm]
                                  \psi_{2N}    &\psi_{2N}^{(1)}  &\cdots &\psi_{2N}^{(N-2)}&\psi_{2N}^{(N-1)}      &\phi_{2N}  &\phi_{2N}^{(1)} &\cdots  &\phi_{2N}^{(N-2)}       &-\psi_{2N}^{(N)}
\end{array}
\right|,
\end{equation}

\begin{equation}\label{427}
\Delta_{C_{N-1}}=\left|
\begin{array}{ccccccccccccccccc}  \psi_1       &\psi_1^{(1)}     &\cdots &\psi_1^{(N-2)}   &-\phi_1^{(N)}         &\phi_1     &\phi_1^{(1)}    &\cdots &\phi_1^{(N-2)}           &\phi_1^{(N-1)}\\[1mm]
                                  \psi_2       &\psi_2^{(1)}     &\cdots &\psi_2^{(N-2)}   &-\phi_2^{(N)}         &\phi_2     &\phi_2^{(1)}    &\cdots  &\phi_2^{(N-2)}          &\phi_2^{(N-1)}\\[1mm]
                                  \vdots       &\vdots           &\ddots &\vdots           &\vdots                 &\vdots     &\vdots          &\ddots  &\vdots                  &\vdots\\[1mm]
                                  \psi_{2N}    &\psi_{2N}^{(1)}  &\cdots &\psi_{2N}^{(N-2)}&-\phi_{2N}^{(N)}      &\phi_{2N}  &\phi_{2N}^{(1)} &\cdots  &\phi_{2N}^{(N-2)}       &\phi_{2N}^{(N-1)}
\end{array}
\right|.
\end{equation}
In Eqs. ({\ref{413})-({\ref{427}), we have used the notations $x[0]=x$,
$u[0]=u$, $v[0]=v$,
 $\psi_k^{(j)}=\lambda_k^{-j}\psi_k$ and
$\phi_k^{(j)}=\lambda_k^{-j}\phi_k$
$(k=1,2,\ldots,2N;\,\,j=1,2,\ldots,N)$.

{\bf Proof 1$^{\prime}$.} Let $D^{-1}=D^{*}/\det D$ and
\begin{equation}\label{428}
(D_y+DU)D^{*}=\left(\begin{array}{cc}  f_{11}(\lambda)    &  f_{12}(\lambda)\\[1mm]
                                                  f_{21}(\lambda)    &  f_{22}(\lambda)
\end{array}
\right).
\end{equation}
It is not hard for us  to see that $f_{11}(\lambda)$, $f_{12}(\lambda)$,
$f_{21}(\lambda)$, and $f_{22}(\lambda)$ are $(-2N+1)$th degree
polynomials in $\lambda$. Substituting (\ref{40}) into the first
equation of (\ref{10}) yields
\begin{equation}\label{429}
\psi_{ky}=\lambda_k(x_y\psi_k+u_y\phi_k),\,\,\,
\phi_{ky}=\lambda_k(v_y\psi_k-x_y\phi_k),\,\,\,(0\leq k\leq 2N).
\end{equation}
A direct calculation reveals that all $\lambda_k (1\leq k\leq 2N)$ are
the roots of $f_{mn} (m,n=1,2)$. Therefore from (\ref{428}), we have

\begin{equation}\label{430}
(D_y+DU)D^{*}=(\det D)\lambda P,
\end{equation}
where
\begin{equation}\label{431}
\lambda P=\lambda\left(\begin{array}{cc}  p^{(1)}_{11}    &  p^{(1)}_{12} \\[1mm]
                                           p^{(1)}_{21}    &  p^{(1)}_{22}
\end{array}
\right)
\end{equation}
and $p^{(i)}_{mn} (m,n=1,2; i=1)$ are independent of $\lambda$. Apparently,
Eq. (\ref{430}) can be rewritten as
\begin{equation}\label{432}
D_y+DU=\lambda PD.
\end{equation}
Comparing the coefficients of $\lambda^{-N+1}$ and
$\lambda^{-N+m}$ on both sides of (\ref{432}) and noticing (\ref{413}), we have
\begin{equation}\label{433}
\begin{array}{ll}
p^{(1)}_{11}=x_y+A_{N-1,y}=x[1]_y,\,\,\,p^{(1)}_{12}=u_y+B_{N-1,y}=u[1]_y\\[2mm]
p^{(1)}_{21}=v_y+C_{N-1,y}=v[1]_y,\,\,\,p^{(1)}_{22}=-x_y+E_{N-1,y}=-x[1]_y,
\end{array}
\end{equation}
\begin{equation}\label{434}
\begin{array}{ll}
A_{N-m,y}=-A_{N-m+1}(x_y-x[1]_y)+C_{N-m+1}u[1]_y-B_{N-m+1}v_y,\\[2mm]
B_{N-m,y}=B_{N-m+1}(x_y+x[1]_y)-A_{N-m+1}u_y+E_{N-m+1}u[1]_y,\\[2mm]
C_{N-m,y}=C_{N-m+1}(x_y+x[1]_y)+A_{N-m+1}v[1]_y-E_{N-m+1}v_y,\\[2mm]
E_{N-m,y}=E_{N-m+1}(x_y-x[1]_y)+B_{N-m+1}v[1]_y-C_{N-m+1}u_y,\,\,(2\leq
m\leq N).
\end{array}
\end{equation}
From Eqs. (\ref{16}) and (\ref{431}), one can readily see that $P= R[1]$. Thus, the proof is
completed.

{\bf Proposition 2$^{\prime}$.} Under the Darboux transformation (\ref{411}), the matrices $S[1]$ and $W[1]$
defined by the second and third equations of (\ref{16}) have the same
form as $S$ and $W$ in the second and third equation of (\ref{12}).

{\bf Proof 2$^{\prime}$.} Let $D^{-1}=D^{*}/\det D$ and
\begin{equation}\label{435}
(D_\tau+DV)D^{*}=\left(\begin{array}{cc}  g_{11}(\lambda)    &  g_{12}(\lambda)\\[1mm]
                                          g_{21}(\lambda)    &  g_{22}(\lambda)
\end{array}
\right).
\end{equation}
One may easily see that $g_{11}(\lambda)$, $g_{12}(\lambda)$,
$g_{21}(\lambda)$, and $g_{22}(\lambda)$ are the $(-2N-1)$th degree
polynomials in $\lambda$. Substituting (\ref{40}) into the second
equation of (\ref{10}) leads to
\begin{equation}\label{436}
\psi_{k\tau}=\dfrac{1}{4\lambda_k}\psi_k-\dfrac{1}{2}u\phi_k,\,\,\,
\phi_{k\tau}=\dfrac{1}{2}v\psi_k-\dfrac{1}{4\lambda_k}\phi_k,\,\,\,(0\leq
k\leq 2N).
\end{equation}
A direct calculation shows that all $\lambda_k (1\leq k\leq 2N)$ are
the roots of $g_{mn} (m,n=1,2)$. Therefore, we arrive at
\begin{equation}\label{437}
(D_\tau+DV)D^{*}=(\det D)Q(\lambda),
\end{equation}
where
\begin{equation}\label{438}
Q(\lambda)=\dfrac{\lambda}{2}\left(\begin{array}{cc}  q^{(0)}_{11}    &  q^{(0)}_{12} \\[1mm]
                                                      q^{(0)}_{21}    &  q^{(0)}_{22}
\end{array}
\right)+\dfrac{1}{4\lambda}\left(\begin{array}{cc}  q^{(1)}_{11}    &  q^{(1)}_{12}\\[1mm]
                                                    q^{(1)}_{21}    &  q^{(1)}_{22}
\end{array}
\right)
\end{equation}
and $q^{(i)}_{mn} (m,n=1,2; i=1,2)$ are independent of $\lambda$.
Hence, Eq. (\ref{437}) can be rewritten as
\begin{equation}\label{439}
D_\tau+DV=Q(\lambda)D.
\end{equation}
Comparing the coefficients of $\lambda^{-N-1}$ and $\lambda^{-N}$ on both sides of Eq.
(\ref{439}), we obtain
\begin{equation}\label{440}
\begin{array}{ll}
q^{(1)}_{11}=-q^{(1)}_{22}=1,\,\,\,q^{(1)}_{21}=q^{(1)}_{12}=0 ,
\end{array}
\end{equation}
\begin{equation}\label{441}
\begin{array}{ll}
q^{(0)}_{21}=v+C_{N-1}=v[1],\,\,\,
q^{(0)}_{12}=-u-B_{N-1}=-u[1],\,\,\,
q^{(0)}_{11}=q^{(0)}_{22}=0.
\end{array}
\end{equation}
From Eqs. (\ref{16}) and (\ref{438}), it is not hard for us to see that $Q(\lambda)=V$, which
completes the proof.

{\bf Remark 1:}

(i) If $(x(y,\tau),\,u(y,\tau),\,v(y,\tau))$ is a
parametric representation of the solution to the 2SP system (\ref{2}), then 
$(x(-y,\tau),\,u(-y,\tau),\,v(-y,\tau))$ is a solution, too; and for
(
an arbitrary complex constant $C$, $(x(y,\tau)+C,\,u(y,\tau),\,v(y,\tau))$
is also a solution;

(ii) The 2SP system (\ref{2}) has a trivial solution $(x,u,v)$=$(\alpha y+\beta
\tau+x_0,0,0)$, where $\alpha,\beta, x_0\in\mathbb{R}$.

Substituting the trivial solution $x=\alpha y+\beta \tau+x_0$,
$u=v=0$ into the Lax system (\ref{10}), we obtain 
\begin{equation}\label{43}
\left(\begin{array}{cc}  \psi_{(1,k)}  \\[1mm]
                         \phi_{(1,k)}
\end{array}
\right)=\left(\begin{array}{cc}  e^{\xi_k}\\[1mm]
                                 0
\end{array}
\right),\,\,\,\,\,\,\left(\begin{array}{cc}  \psi_{(2,k)}  \\[1mm]
                                             \phi_{(2,k)}
\end{array}
\right)=\left(\begin{array}{cc}  0\\[1mm]
                                 e^{-\xi_k}
\end{array}
\right),
\end{equation}
where $\xi_k=\alpha\lambda_k y+\frac{1}{4\lambda_k} \tau,\,\, (1\leq
k \leq 2N)$. Thus as per (\ref{40}), we accordingly have 
\begin{equation}\label{44}
\left(\begin{array}{cc}  \psi_k  \\[1mm]
                         \phi_k
\end{array}
\right)=\left(\begin{array}{cc} \mu_k e^{\xi_k} \\[1mm]
                                e^{-\xi_k}
\end{array}
\right).
\end{equation}

Under the condition $\lambda_{2k-1}\,\lambda_{2k}<0$,\,\,$\mu_{2k-1}\mu_{2k}<0$
$(k=1,2,\cdots,N)$, substituting (\ref{44}) into (\ref{413}) generates
the $N$-soliton
solution of the 2SP system (\ref{2}).

{\bf Remarks 2:} If $\lambda_{2k}=-\lambda_{2k-1}$
and $\mu_{2k}\,\mu_{2k-1}=-1$ $(k=1,2,\cdots,N)$, (\ref{413}) and (\ref{44})
can be reduced to the $N$-soliton solution of the short pulse
equation (\ref{1}).

Let us take three special cases $N=1$, $N=2$
and $N=3$ as examples.

{\bf Case 1 ($N=1$).} \,\,From (\ref{413}), we have the following one-soliton
solution of the 2SP system (\ref{2}) 
\begin{eqnarray}\label{45}
x[1]&=&\alpha y+\beta \tau+x_0-\dfrac{\lambda_2\mu_1{\rm
e}^{\zeta_1}-\lambda_1\mu_2 {\rm
e}^{\zeta_2}}{\lambda_1\lambda_2(\mu_1{\rm e}^{\zeta_1}-\mu_2 {\rm
e}^{\zeta_2})},\\
\label{46}
u[1]&=&u[0]-\dfrac{(\lambda_1-\lambda_2)\mu_1 \mu_2 {\rm
e}^{\zeta_1+\zeta_2}}{\lambda_1\lambda_2(\mu_1{\rm
e}^{\zeta_1}-\mu_2 {\rm
e}^{\zeta_2})},\\
\label{46r}
v[1]&=&v[0]+\dfrac{\lambda_1-\lambda_2}{\lambda_1\lambda_2(\mu_1{\rm
e}^{\zeta_1}-\mu_2 {\rm e}^{\zeta_2})},
\end{eqnarray}
where   $\zeta_k=2(\alpha \lambda_k y+\dfrac{1}{4\lambda_k}\tau)$
$(k=1,2)$, $u[0]=u=0$, $v[0]=v=0$ and $x_0=0$, and parameters 
satisfy $\lambda_1\lambda_2<0$,\,\,$\mu_1\mu_2<0$.

According to Remark 2, if we substitute $\lambda_2=-\lambda_1$,
$\mu_2=-\mu_1^{-1}$  into (\ref{45})-(\ref{46}), then Eqs. (\ref{46}) and  (\ref{46r}) read 
one-loop soliton solution of the short pulse equation (\ref{1}) as follows
\begin{eqnarray}
\label{47}
x[1]&=&\alpha y+\beta
\tau-\dfrac{1}{\lambda_1}\tanh(\zeta_1+\ln|\mu_1|),\\
\label{48}
u[1]=v[1]&=&-\dfrac{1}{\lambda_1}\sech(\zeta_1+\ln|\mu_1|),
\end{eqnarray}
where $\zeta_1=2(\alpha \lambda_1 y+\dfrac{1}{4\lambda_1}\tau)$.

\begin{center}
\resizebox{2.5in}{!}{\includegraphics{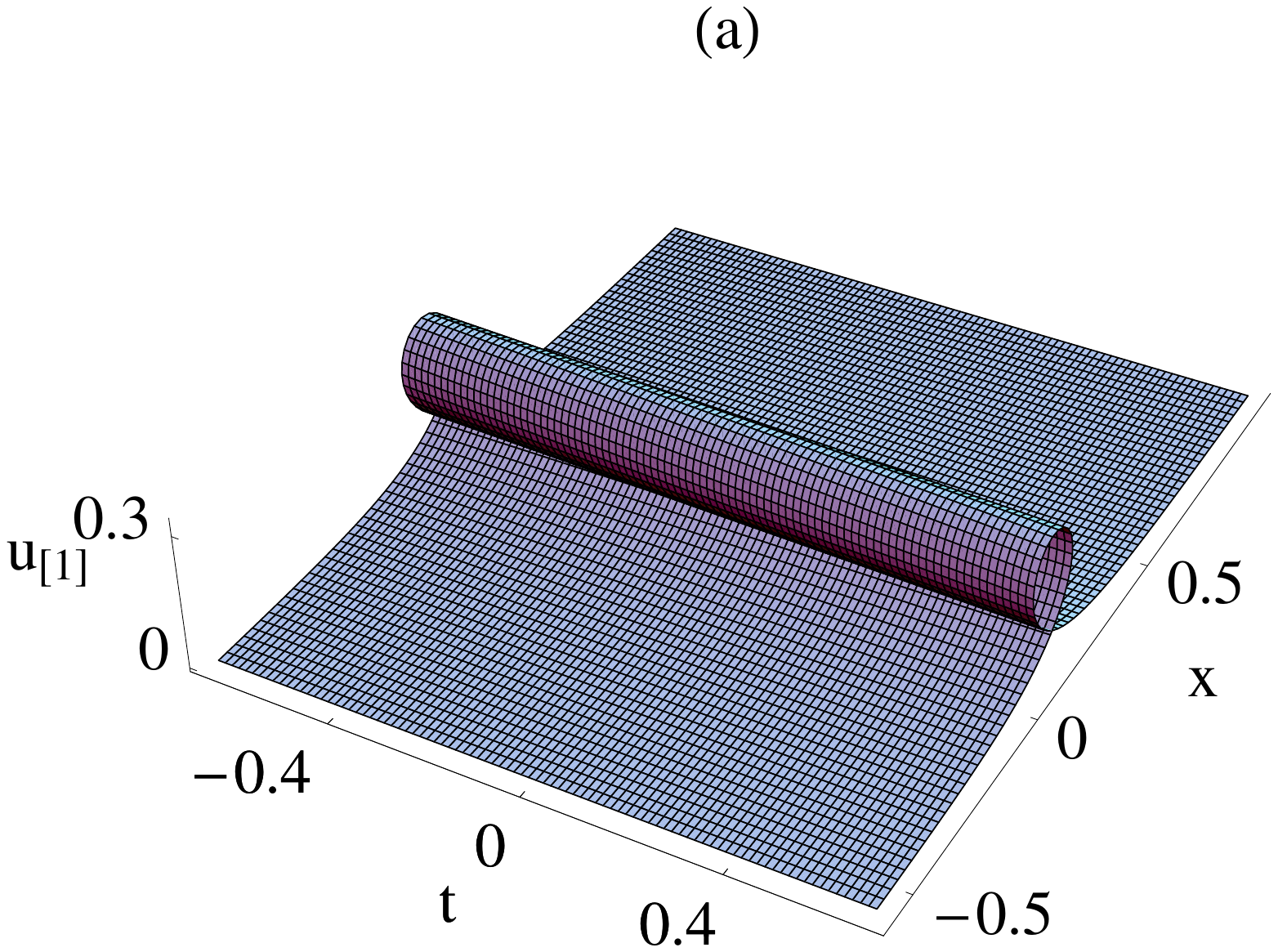}}
\resizebox{2.5in}{!}{\includegraphics{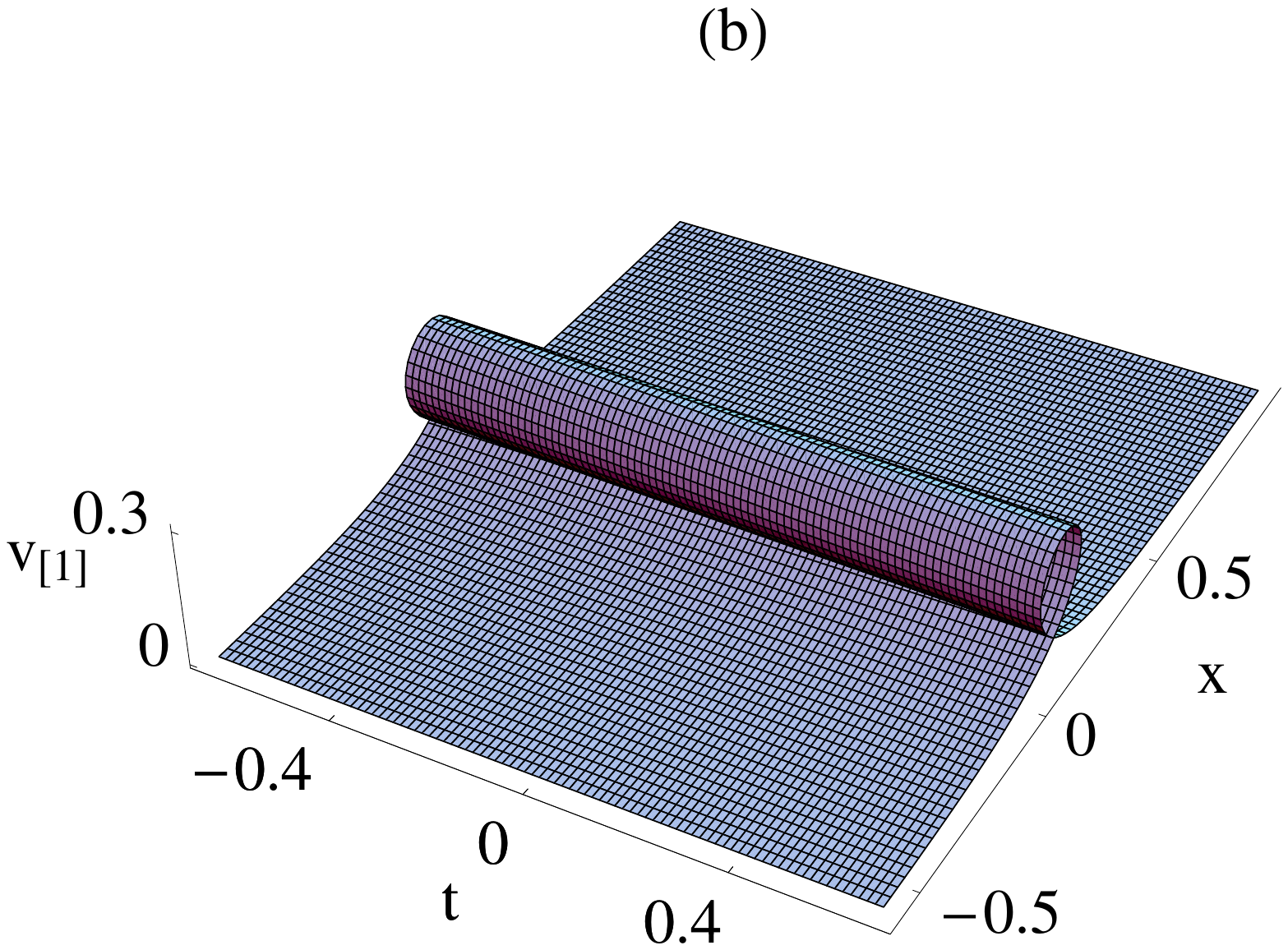}}
\end{center}
\centerline{\small{Figure 1. \,The time evolution of one-loop
soliton solution (\ref{45})-(\ref{46}) with $\lambda_1=-3$,}}
\centerline{\small{$\lambda_2=\frac{10}{3}$,\,
$\mu_1=2,\,\,\mu_2=-\frac{1}{2}$,\,\,$\alpha=1$,
 and $\beta=0$.}}

\begin{center}
\resizebox{2.5in}{!}{\includegraphics{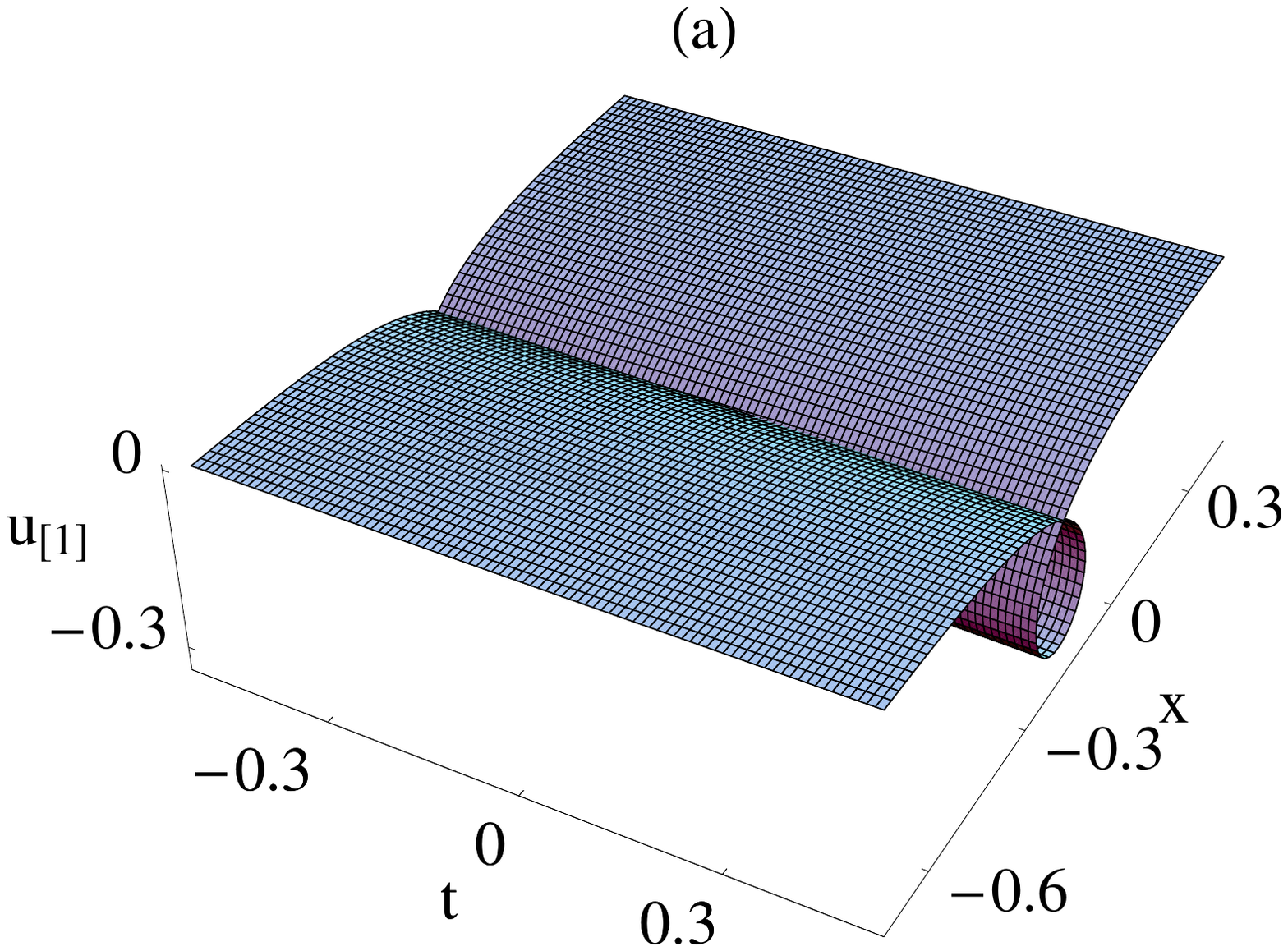}}
\resizebox{2.5in}{!}{\includegraphics{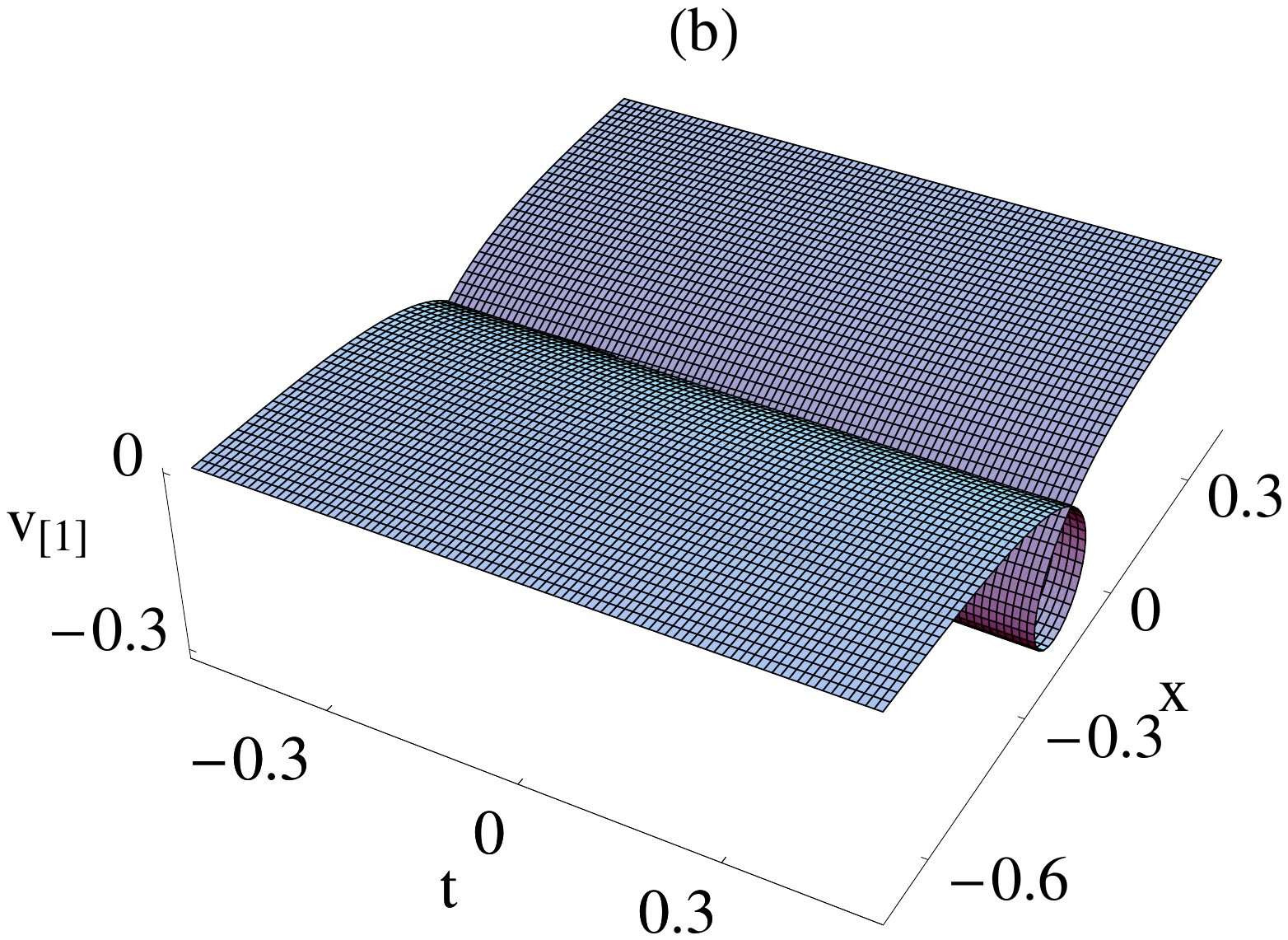}}
\end{center}
\centerline{\small{Figure 2. \,The time evolution of one-anti-loop
soliton solution (\ref{45})-(\ref{46}) with $\lambda_1=3$,}}
\centerline{\small{$\lambda_2=-\frac{10}{3}$,\,$\mu_1=2,\,\,\mu_2=-\frac{1}{2}$,\,\,$\alpha=1$,
 and $\beta=0$.}}

Choosing appropriately different parameters may send
(\ref{45})-(\ref{46}) to different types of soliton
solutions for the 2SP system (\ref{2}). For instance,
the time evolutions of one-loop soliton and one-anti-loop soliton
solutions are captured and illustrated in Figure 1 and Figure 2, respectively.

{\bf Case 2 ($N=2$).} \,\, From (\ref{413}), we may obtain a two-soliton
solution of the 2SP system (\ref{2}) in the following form
\begin{equation}\label{49}
x[2]=\alpha y+\beta
\tau+x_0+\dfrac{\Delta_{A_1}}{\Delta_{1}},\,\,\,\,u[2]=u[0]+\dfrac{\Delta_{B_1}}{\Delta_{1}},\,\,\,
v[2]=v[0]+\dfrac{\Delta_{C_1}}{\Delta_{1}},
\end{equation}
where
\begin{equation}\label{50}
\Delta_{1}=\left|
\begin{array}{cccccccc}  \psi_1         &\psi_1^{(1)}   &\phi_1     &\phi_1^{(1)}\\[1mm]
                         \psi_2         &\psi_2^{(1)}   &\phi_2     &\phi_2^{(1)}\\[1mm]
                         \psi_3         &\psi_3^{(1)}   &\phi_3     &\phi_3^{(1)}\\[1mm]
                         \psi_4         &\psi_4^{(1)}   &\phi_4     &\phi_4^{(1)}
\end{array}
\right|,\,\,\,\Delta_{A_1}=\left|
\begin{array}{cccccccc}  \psi_1         &-\psi_1^{(2)}   &\phi_1     &\phi_1^{(1)}\\[1mm]
                         \psi_2         &-\psi_2^{(2)}   &\phi_2     &\phi_2^{(1)}\\[1mm]
                         \psi_3         &-\psi_3^{(2)}   &\phi_3     &\phi_3^{(1)}\\[1mm]
                         \psi_4         &-\psi_4^{(2)}   &\phi_4     &\phi_4^{(1)}
\end{array}
\right|,
\end{equation}
\begin{equation}\label{51}
\Delta_{B_1}=\left|
\begin{array}{cccccccc}  \psi_1         &\psi_1^{(1)}   &\phi_1     &-\psi_1^{(2)}\\[1mm]
                         \psi_2         &\psi_2^{(1)}   &\phi_2     &-\psi_2^{(2)}\\[1mm]
                         \psi_3         &\psi_3^{(1)}   &\phi_3     &-\psi_3^{(2)}\\[1mm]
                         \psi_4         &\psi_4^{(1)}   &\phi_4     &-\psi_4^{(2)}
\end{array}
\right|,\,\,\,\Delta_{C_1}=\left|
\begin{array}{cccccccc}  \psi_1         &-\phi_1^{(2)}   &\phi_1     &\phi_1^{(1)}\\[1mm]
                         \psi_2         &-\phi_2^{(2)}   &\phi_2     &\phi_2^{(1)}\\[1mm]
                         \psi_3         &-\phi_3^{(2)}   &\phi_3     &\phi_3^{(1)}\\[1mm]
                         \psi_4         &-\phi_4^{(2)}   &\phi_4     &\phi_4^{(1)}
\end{array}
\right|,
\end{equation}
$u[0]=v[0]=x_0=0$,  $\psi_k^{(j)}=\lambda_k^{-j}\psi_k$,
$\phi_k^{(j)}=\lambda_k^{-j}\phi_k$ $(k=1,2,3,4;\,\,j=1,2)$, and $(\psi_k,\phi_k)^{T}$
are given through (\ref{44}).
The two-soliton solution
(\ref{49}) has interactional dynamics, for  instance, two-loop soliton with
$\lambda_1=-\lambda_2=3,\,\lambda_3=-\lambda_4=-2$,\,$\mu_1=2,\,\,\mu_2=-\frac{1}{2}$,\,\,
$\mu_3=3$,\,\,$\mu_4=-1$,\,$\alpha=1$,\,$\beta=0$ (see figure 3),
two-anti-loop soliton with
 $\lambda_1=-\lambda_2=-3,\,\,\lambda_3=-\lambda_4=2$,\,$\mu_1=2,\,\,\mu_2=-\frac{1}{2}$,\,\,
$\mu_3=3$,\,\,$\mu_4=-1$,\,\,$\alpha=1$,\, $\beta=0$ (see figure 4),
and loop-anti-loop soliton with
 $\lambda_1=-\lambda_2=-3,\,\,\lambda_3=-\lambda_4=-5$,\,$\mu_1=2,\,\,\mu_2=-\frac{1}{2}$,\,\,
$\mu_3=3$,\,\,$\mu_4=-1$,\,\,$\alpha=1$,\, $\beta=0$ (see figure 5).

The following Figures 3-5 show  the dynamical interaction of two-soliton solutions $(u[2],v[2])$ at three different times: (a) the short dashed line 
stands for the wave
elevation at $t=-30$, (b) the long dashed line 
represents for the wave
elevation at $t=0$, and (c) the solid line 
refers to
the wave
elevation at $t=30$. Initially,  two-solitons are well
separated at $t=-30$ and the soliton with faster velocity is located at the
right. At $t=0$, the soliton with faster velocity catches up with
the other soliton with slower velocity and they have a collision.
Afterward, they gradually separate as time increases and eventually
at $t=30$, the two solitons recover their original shapes apart from
some phase shifts. Apparently, 
such a collision is elastic, and
there is no change in shape and amplitude of solitons except a phase
shift. This interaction is very similar to that of two solitons of
the Korteweg-de Vries (KdV) equation. However, an interesting
phenomenon is that   a soliton (see $v[2]$ in Figures 3-4) with
a smaller amplitude could travel faster than the one with larger
amplitude when they interact.
\begin{center}
\resizebox{2.5in}{!}{\includegraphics{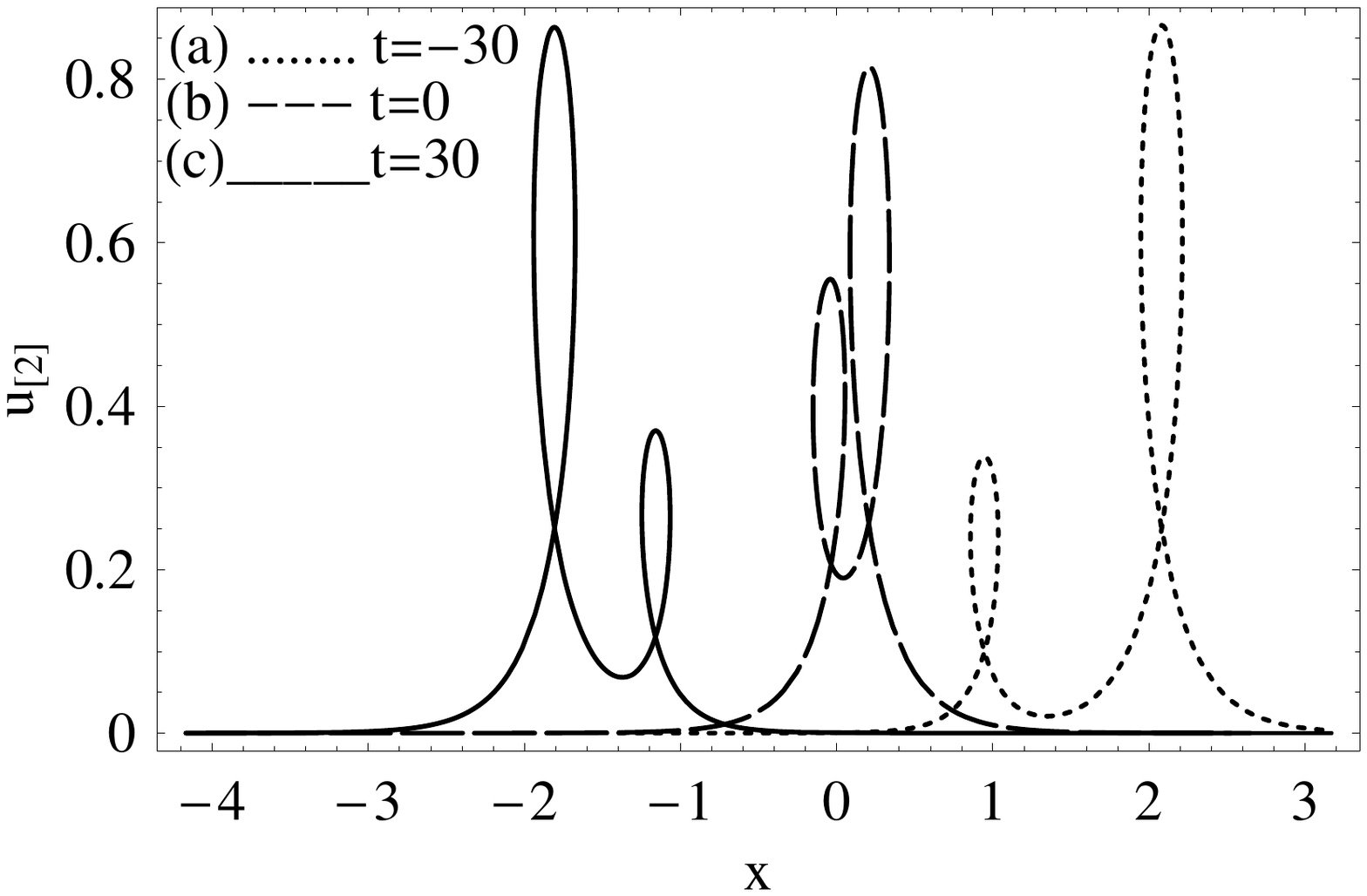}}
\resizebox{2.5in}{!}{\includegraphics{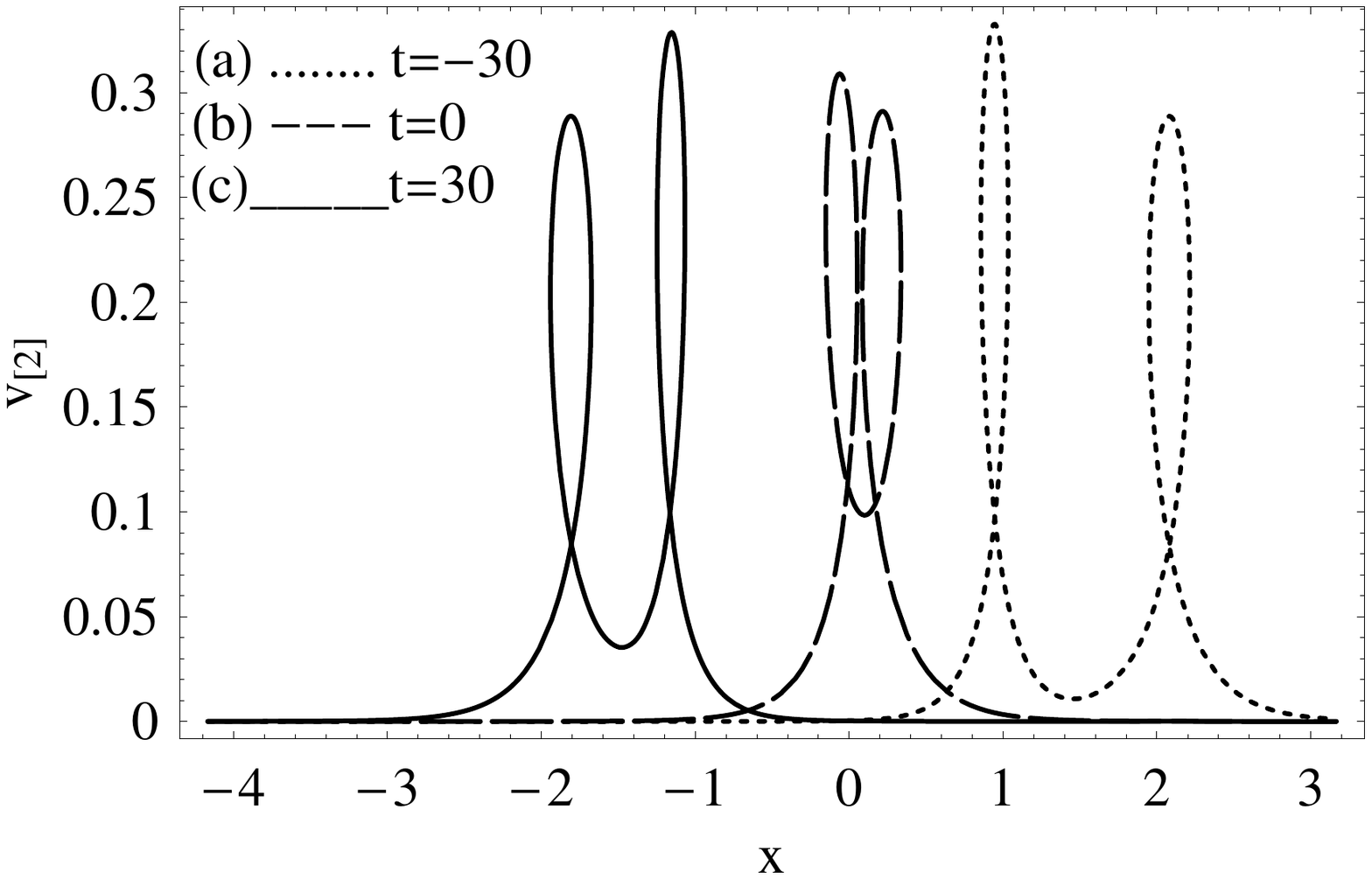}}
\end{center}
\centerline{\small{Figure 3. \,The dynamical interaction of two-loop soliton
solution (\ref{49}) with
$\lambda_1=-\lambda_2=3,\,\,\lambda_3=-\lambda_4=-2$,}}
\centerline{\small{$\mu_1=2,\,\,\mu_2=-\frac{1}{2}$,\,\,
$\mu_3=3$,\,\,$\mu_4=-1$,\,\,$\alpha=1$,
 and $\beta=0$.}}

\begin{center}
\resizebox{2.5in}{!}{\includegraphics{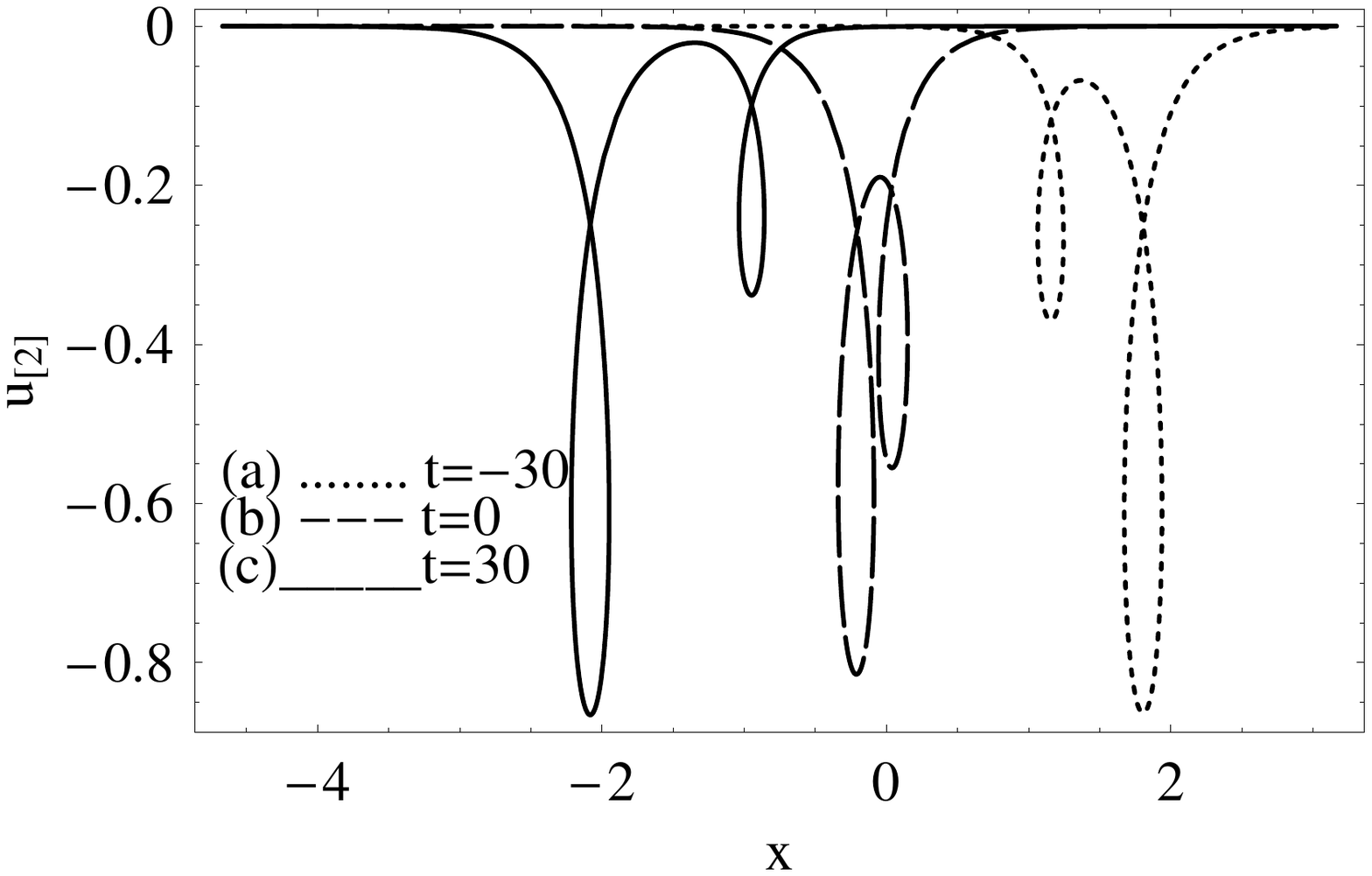}}
\resizebox{2.5in}{!}{\includegraphics{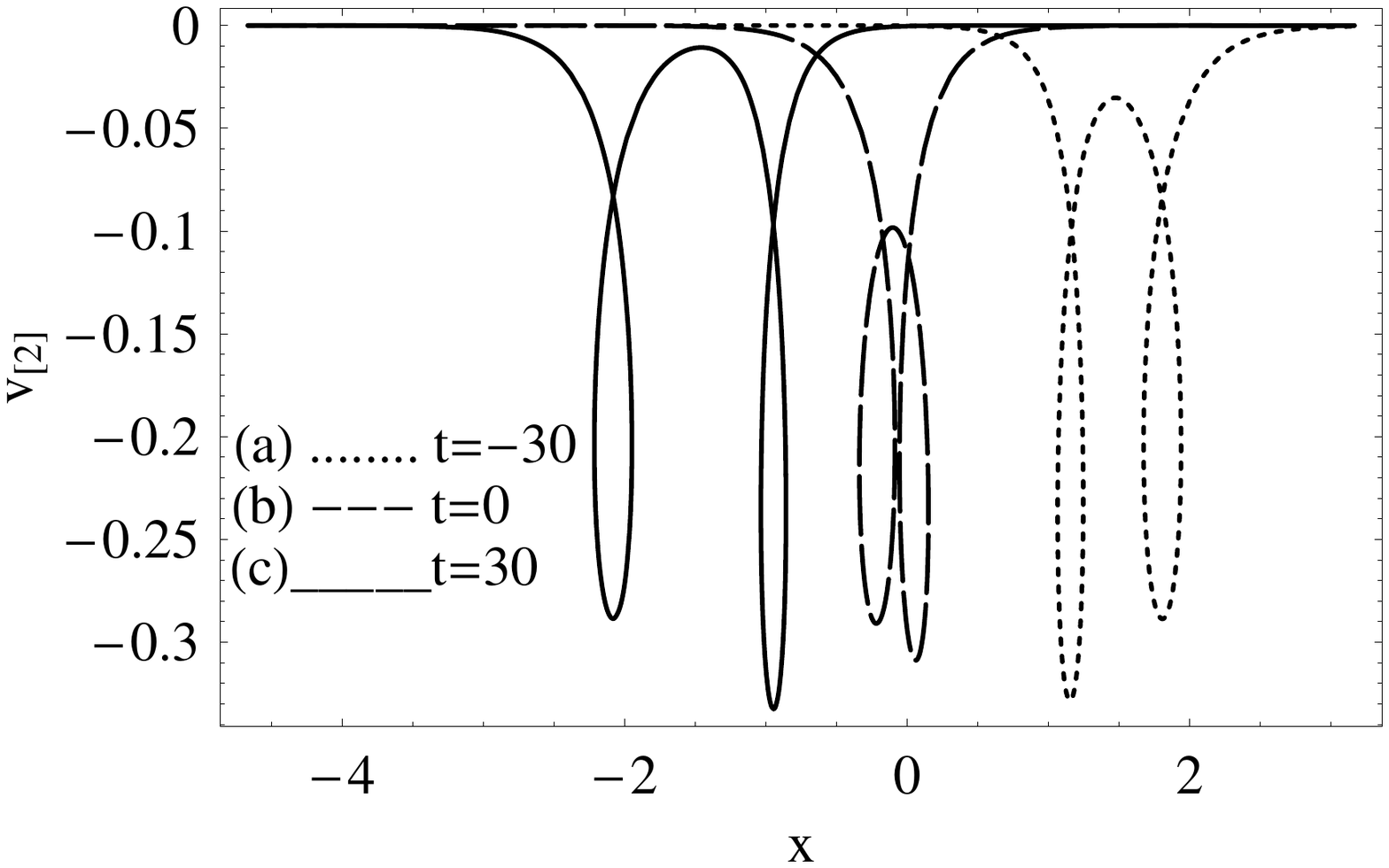}}
\end{center}
\centerline{\small{Figure 4. \,The dynamical interaction of two-anti-loop
soliton solution (\ref{49}) with
$\lambda_1=-\lambda_2=-3,\,\,\lambda_3=-\lambda_4=2$,}}
\centerline{\small{$\mu_1=2,\,\,\mu_2=-\frac{1}{2}$,\,\,
$\mu_3=3$,\,\,$\mu_4=-1$,\,\,$\alpha=1$,
 and $\beta=0$.}}

\begin{center}
\resizebox{2.5in}{!}{\includegraphics{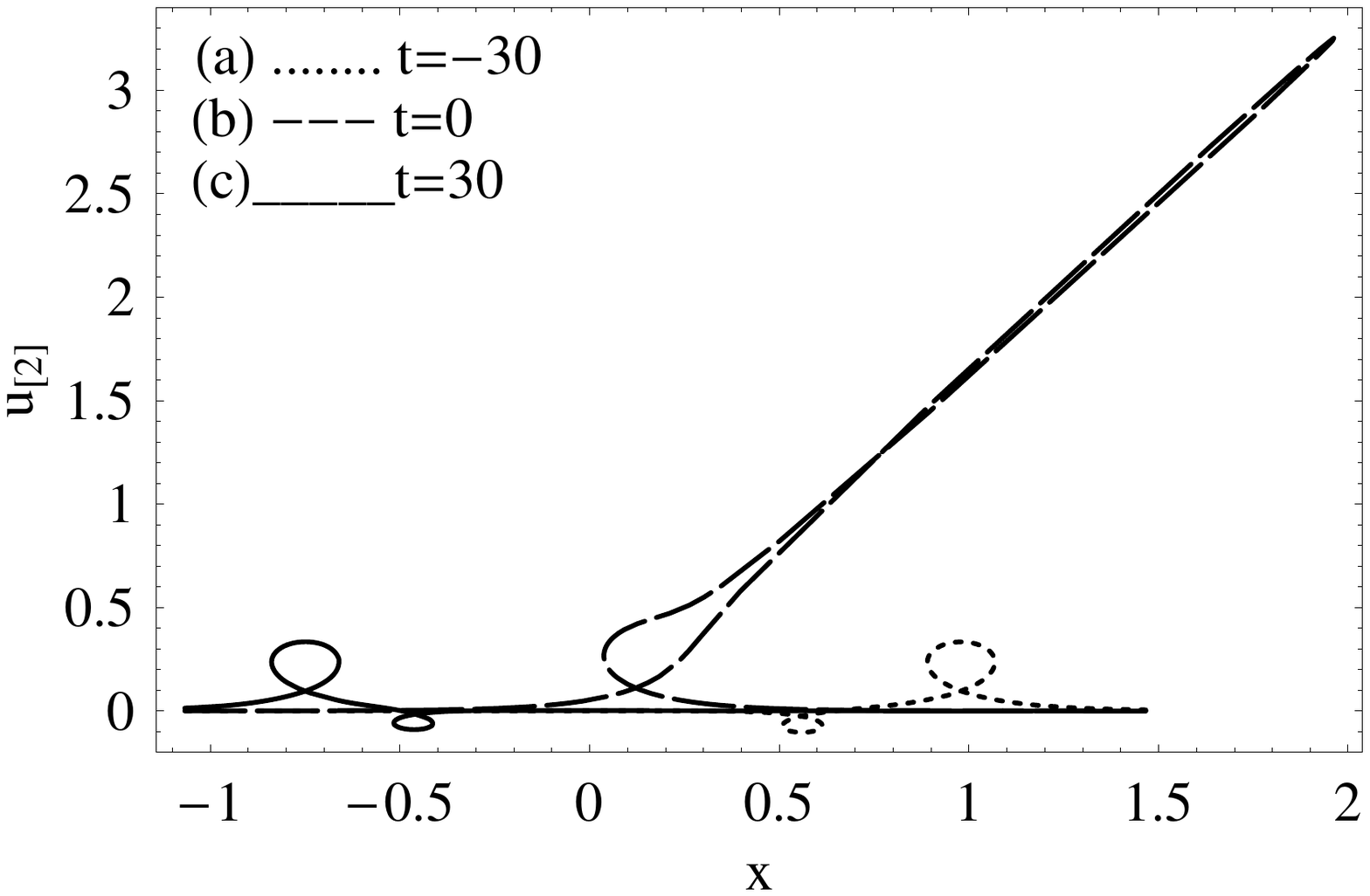}}
\resizebox{2.5in}{!}{\includegraphics{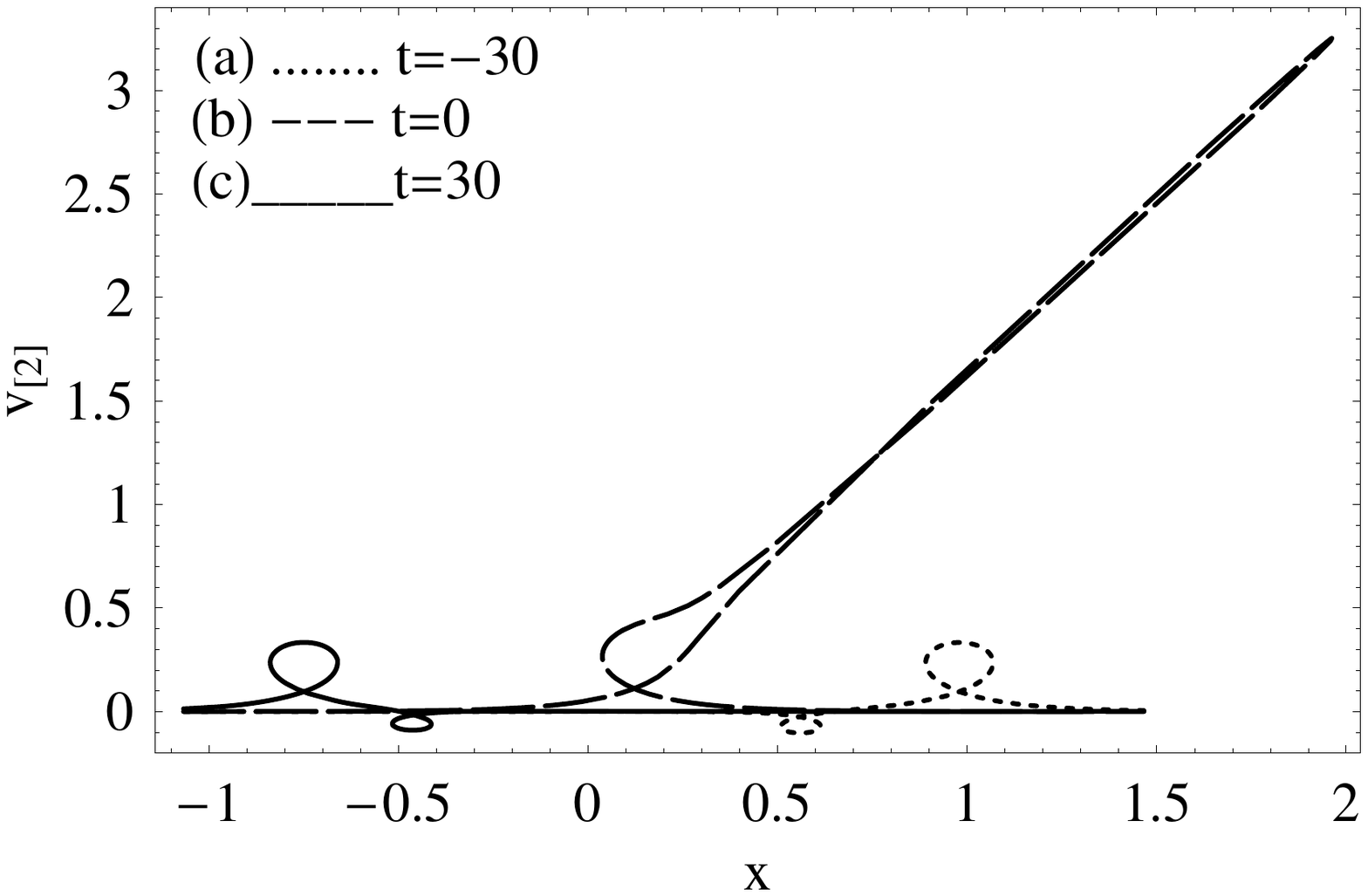}}
\end{center}
\centerline{\small{Figure 5. \,The dynamical interaction of loop-anti-loop
soliton solution (\ref{49}) with
$\lambda_1=-\lambda_2=-3,\,\,\lambda_3=-\lambda_4=-5$,}}
\centerline{\small{$\mu_1=2,\,\,\mu_2=-\frac{1}{2}$,\,\,
$\mu_3=3$,\,\,$\mu_4=-1$,\,\,$\alpha=1$,
 and $\beta=0$.}}

Casting $\lambda_1=\lambda_{1R}+{\rm i}\lambda_{1I}$,\,
$\lambda_2=-(\lambda_{1R}-{\rm
i}\lambda_{1I})$,\,$\lambda_3=\lambda_1^{*}$,\,$\lambda_4=\lambda_2^{*}$
and $\mu_2=-\mu_3=\mu_4=-{\mu_1}$ in Eq. (\ref{49}) produces the following
breather soliton solution of the 2SP system 
(\ref{2}) 
\begin{eqnarray}\label{53}
x[2]&=&\alpha y+\beta
\tau+\dfrac{2\lambda_{1I}\lambda_{1R}}{|\lambda_1|^2}\dfrac{\lambda_{1I}\sinh\Omega_1+
\lambda_{1R}\sin\theta_1}{2\lambda^2_{1I}\cosh^2\Omega_1+\lambda^2_{1R}(1-\cos\theta_1)},\\
\label{54}
u[2]&=&-\dfrac{4\lambda_{1I}\lambda_{1R}\mu_1}{|\lambda_1|^2}\dfrac{\lambda_{1I}\cos\frac{1}{2}\theta_1\cosh\frac{1}{2}\Omega_1
-\lambda_{1R}\sin\frac{1}{2}\theta_1\sinh\frac{1}{2}\Omega_1}{2\lambda^2_{1I}\cosh^2\frac{1}{2}\Omega_1+\lambda^2_{1R}(1-\cos\theta_1)},\\
\label{540}
v[2]&=&-\dfrac{4\lambda_{1I}\lambda_{1R}}{\mu_1|\lambda_1|^2}\dfrac{\lambda_{1I}\cos\frac{1}{2}\theta_1\cosh\frac{1}{2}\Omega_1
+\lambda_{1R}\sin\frac{1}{2}\theta_1\sinh\frac{1}{2}\Omega_1}{2\lambda^2_{1I}\cosh^2\frac{1}{2}\Omega_1+\lambda^2_{1R}(1-\cos\theta_1)},
\end{eqnarray}
where
$\Omega_1=4\alpha\lambda_{1R}y+\frac{\lambda_{1R}}{|\lambda_1|^2}\tau$,\,\,$\theta_1=-4\alpha\lambda_{1I}y+\frac{\lambda_{1I}}{|\lambda_1|^2}\tau$.
The following figure (Figure 6) shows the time evolution of the breather
soliton solutions (\ref{53})-(\ref{540}) with the parameters 
$\lambda_{1R}=\lambda_{1I}=1$,\,$\mu_1=\mu_3=-\mu_2=-\mu_4=\frac{1}{2}$,\,$\alpha=1$,\,
$\beta=0$.

\begin{center}
\resizebox{2.5in}{!}{\includegraphics{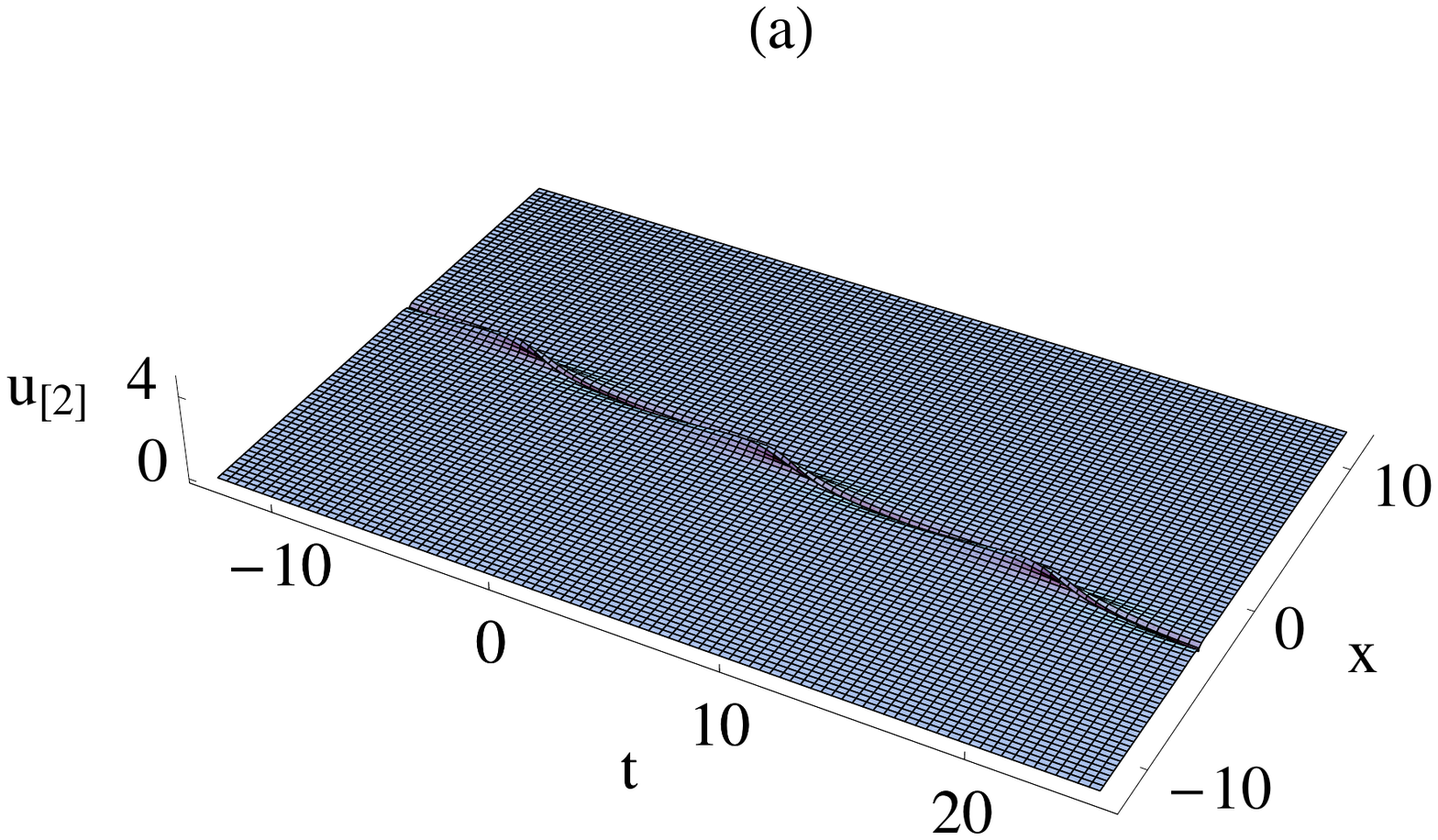}}
\resizebox{2.5in}{!}{\includegraphics{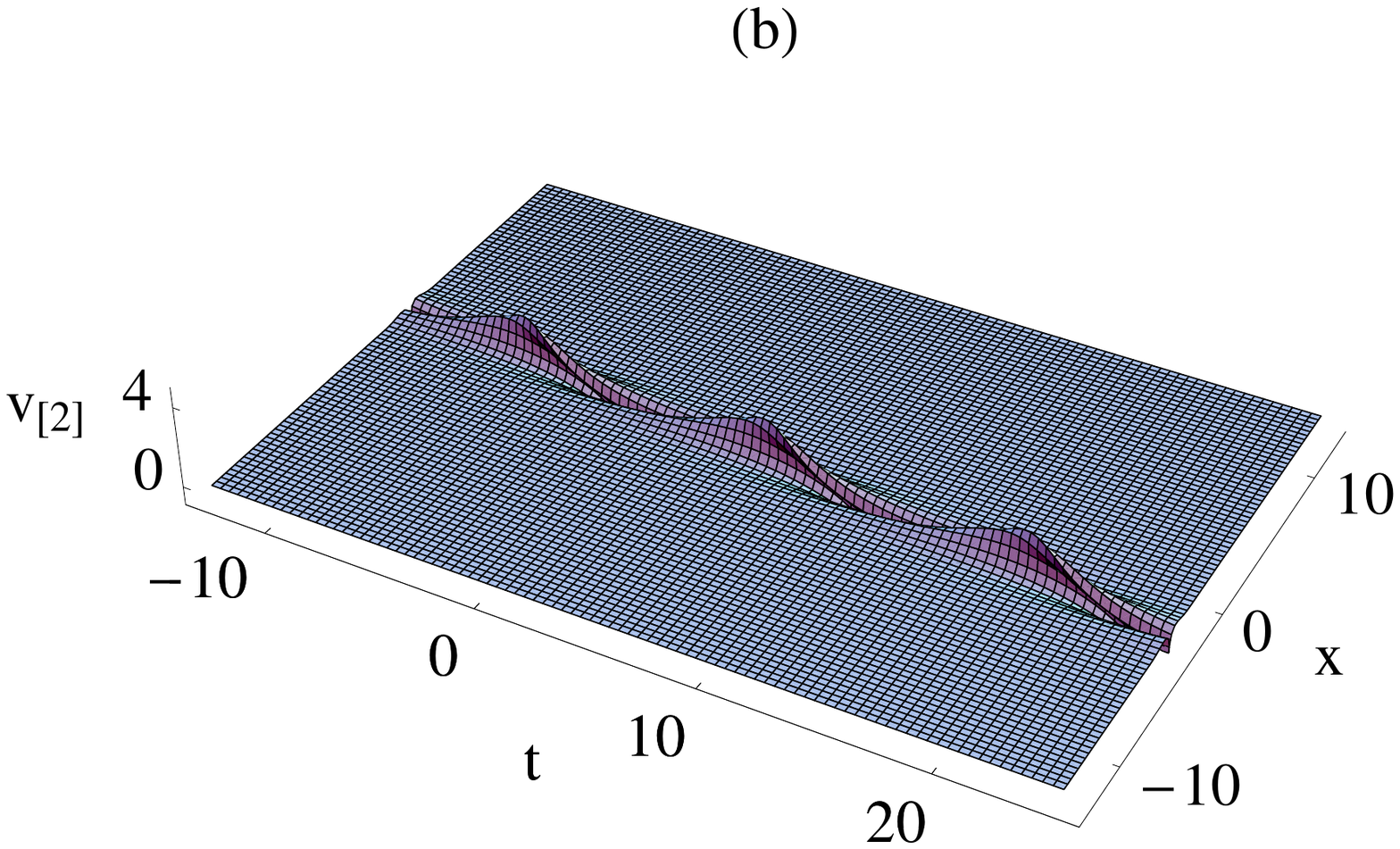}}
\end{center}
\centerline{\small{Figure 6. \,The dynamical interaction of breather-soliton
solutions (\ref{53})-(\ref{54}) with
$\lambda_1=\lambda^{*}_2=-\lambda_3=-\lambda^{*}_4=1+{\rm i}$,}}
\centerline{\small{$\mu_1=2,\,\,\mu_2=-\frac{1}{2}$,\,\,
$\mu_3=3$,\,\,$\mu_4=-1$,\,\,$\alpha=1$,
 and $\beta=0$.}}

 {\bf Case 3 ($N=3$).} \,\, AS per Eq. (\ref{413}), putting $N=3$ generates three-soliton solutions of the 2SP system 
(\ref{2}) 
\begin{equation}\label{55}
x[3]=\alpha y+\beta
\tau+x_0+\dfrac{\Delta_{A_2}}{\Delta_{2}},\,\,\,\,u[3]=u[0]+\dfrac{\Delta_{B_2}}{\Delta_{2}},\,\,\,
v[3]=v[0]+\dfrac{\Delta_{C_2}}{\Delta_{2}},
\end{equation}
where
\begin{equation}\label{56}
\Delta_{2}=\left|
\begin{array}{cccccccc}  \psi_1       &\psi_1^{(1)}   &\psi_1^{(2)}     &\phi_1      &\phi_1^{(1)}   &\phi_1^{(2)}\\[1mm]
                         \psi_2       &\psi_2^{(1)}   &\psi_2^{(2)}     &\phi_2      &\phi_2^{(1)}   &\phi_2^{(2)}\\[1mm]
                         \psi_3       &\psi_3^{(1)}   &\psi_3^{(2)}     &\phi_3      &\phi_3^{(1)}   &\phi_3^{(2)}\\[1mm]
                         \psi_4       &\psi_4^{(1)}   &\psi_4^{(2)}     &\phi_4      &\phi_4^{(1)}   &\phi_4^{(2)}\\[1mm]
                         \psi_5       &\psi_5^{(1)}   &\psi_5^{(2)}     &\phi_5      &\phi_5^{(1)}   &\phi_5^{(2)}\\[1mm]
                         \psi_6       &\psi_6^{(1)}   &\psi_6^{(2)}     &\phi_6      &\phi_6^{(1)}   &\phi_6^{(2)}
\end{array}
\right|,
\end{equation}

\begin{equation}\label{57}
\Delta_{A_2}=\left|
\begin{array}{cccccccc}  \psi_1       &\psi_1^{(1)}   &-\psi_1^{(3)}     &\phi_1      &\phi_1^{(1)}   &\phi_1^{(2)}\\[1mm]
                         \psi_2       &\psi_2^{(1)}   &-\psi_2^{(3)}     &\phi_2      &\phi_2^{(1)}   &\phi_2^{(2)}\\[1mm]
                         \psi_3       &\psi_3^{(1)}   &-\psi_3^{(3)}     &\phi_3      &\phi_3^{(1)}   &\phi_3^{(2)}\\[1mm]
                         \psi_4       &\psi_4^{(1)}   &-\psi_4^{(3)}     &\phi_4      &\phi_4^{(1)}   &\phi_4^{(2)}\\[1mm]
                         \psi_5       &\psi_5^{(1)}   &-\psi_5^{(3)}     &\phi_5      &\phi_5^{(1)}   &\phi_5^{(2)}\\[1mm]
                         \psi_6       &\psi_6^{(1)}   &-\psi_6^{(3)}     &\phi_6      &\phi_6^{(1)}   &\phi_6^{(2)}
\end{array}
\right|,
\end{equation}

\begin{equation}\label{58}
\Delta_{B_2}=\left|
\begin{array}{cccccccc}  \psi_1       &\psi_1^{(1)}   &\psi_1^{(2)}     &\phi_1      &\phi_1^{(1)}   &-\psi_1^{(3)} \\[1mm]
                         \psi_2       &\psi_2^{(1)}   &\psi_2^{(2)}     &\phi_2      &\phi_2^{(1)}   &-\psi_2^{(3)} \\[1mm]
                         \psi_3       &\psi_3^{(1)}   &\psi_3^{(2)}     &\phi_3      &\phi_3^{(1)}   &-\psi_3^{(3)} \\[1mm]
                         \psi_4       &\psi_4^{(1)}   &\psi_4^{(2)}     &\phi_4      &\phi_4^{(1)}   &-\psi_4^{(3)} \\[1mm]
                         \psi_5       &\psi_5^{(1)}   &\psi_5^{(2)}     &\phi_5      &\phi_5^{(1)}   &-\psi_5^{(3)} \\[1mm]
                         \psi_6       &\psi_6^{(1)}   &\psi_6^{(2)}     &\phi_6      &\phi_6^{(1)}   &-\psi_6^{(3)}
\end{array}
\right|,
\end{equation}

\begin{equation}\label{59}
\Delta_{C_2}=\left|
\begin{array}{cccccccc}  \psi_1       &\psi_1^{(1)}   &-\phi_1^{(3)}     &\phi_1      &\phi_1^{(1)}   &\phi_1^{(2)}\\[1mm]
                         \psi_2       &\psi_2^{(1)}   &-\phi_2^{(3)}     &\phi_2      &\phi_2^{(1)}   &\phi_2^{(2)}\\[1mm]
                         \psi_3       &\psi_3^{(1)}   &-\phi_3^{(3)}     &\phi_3      &\phi_3^{(1)}   &\phi_3^{(2)}\\[1mm]
                         \psi_4       &\psi_4^{(1)}   &-\phi_4^{(3)}     &\phi_4      &\phi_4^{(1)}   &\phi_4^{(2)}\\[1mm]
                         \psi_5       &\psi_5^{(1)}   &-\phi_5^{(3)}     &\phi_5      &\phi_5^{(1)}   &\phi_5^{(2)}\\[1mm]
                         \psi_6       &\psi_6^{(1)}   &-\phi_6^{(3)}     &\phi_6      &\phi_6^{(1)}   &\phi_6^{(2)}
\end{array}
\right|,
\end{equation}
where $u[0]=v[0]=x_0=0$, $\psi_k^{(j)}=\lambda_k^{-j}\psi_k$,
$\phi_k^{(j)}=\lambda_k^{-j}\phi_k$ $(k=1,2,3,4,5,6;\,\,j=1,2,3)$ and $(\psi_k,\phi_k)^{T}$ are
 given by (\ref{44}).

The following graphs (Figures 7-9) describe the dynamical interactions of the
three-loop-soliton solutions (\ref{55}) with those designated parameters.
One can easily see that the parameters have
little influence on the profiles and characters of the three-soliton solutions of
$(u[3],v[3])$. All the interactional dynamics is very similar to the case of $N=2$.

\begin{center}
\resizebox{2.5in}{!}{\includegraphics{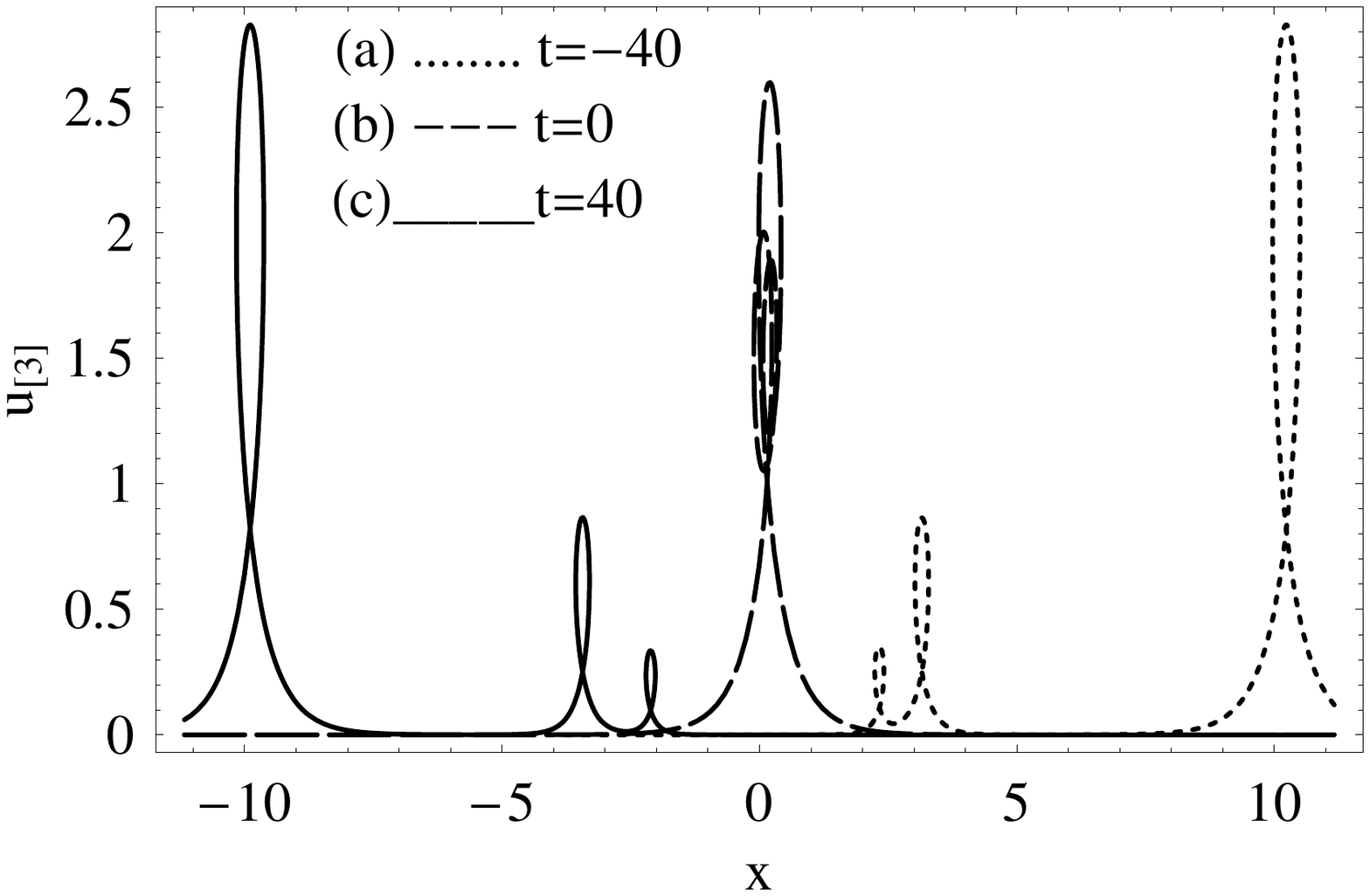}}
\resizebox{2.5in}{!}{\includegraphics{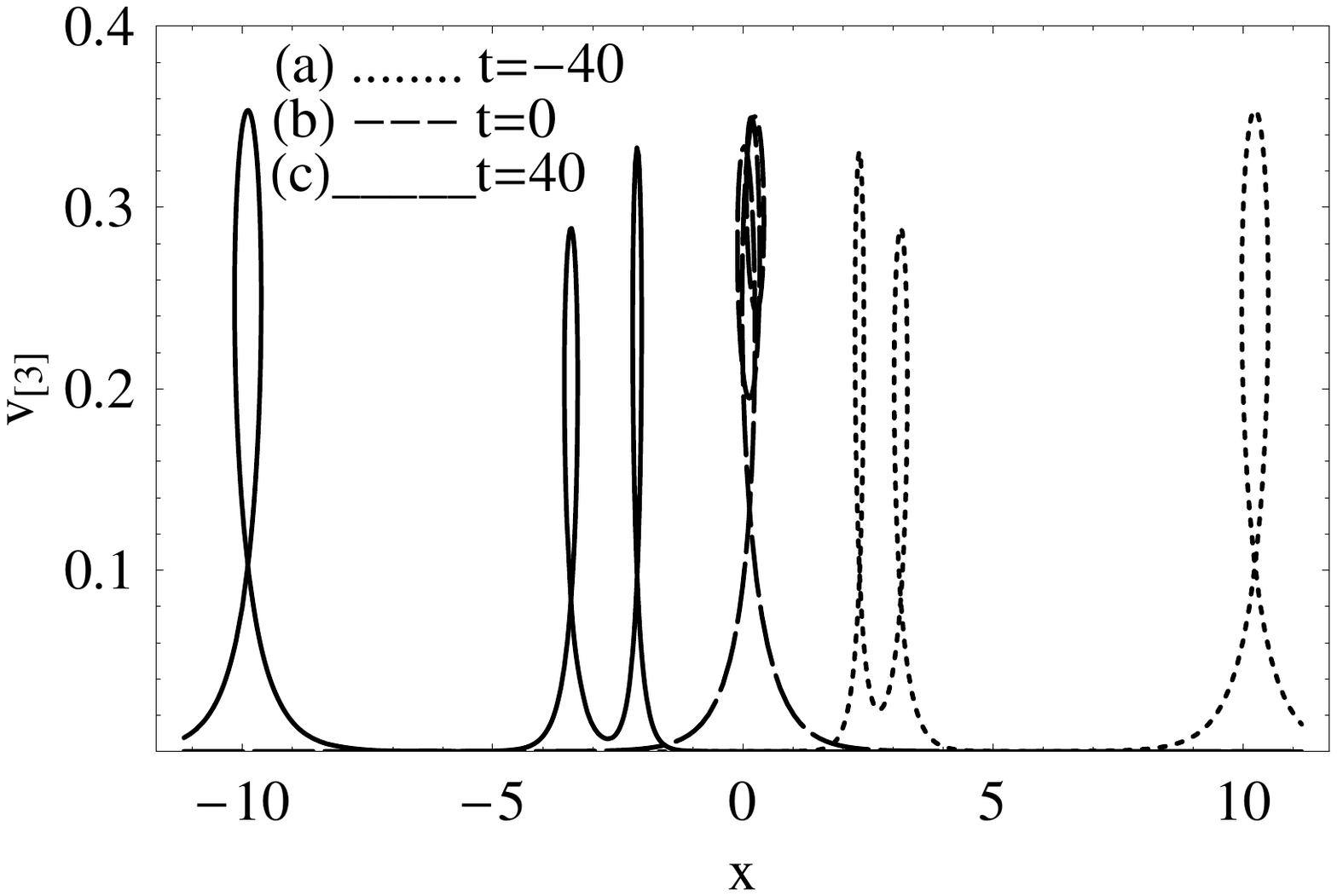}}
\end{center}
\centerline{\small{Figure 7. \,The interactional dynamics of three-soliton
solution (\ref{55}) with
$\lambda_1=-\lambda_2=-3$,\,\,$\lambda_3=-\lambda_4=2$,\,\,$\lambda_5=-\lambda_6=-1$}}
\centerline{\small{$\mu_1=2,\,\,\mu_2=-\frac{1}{2}$,\,\,
$\mu_3=3$,\,\,$\mu_4=-1$,\,\,$\mu_5=4$,\,\,$\mu_6=-2$,$\alpha=1$,
 and $\beta=0$.}}
\begin{center}
\resizebox{2.5in}{!}{\includegraphics{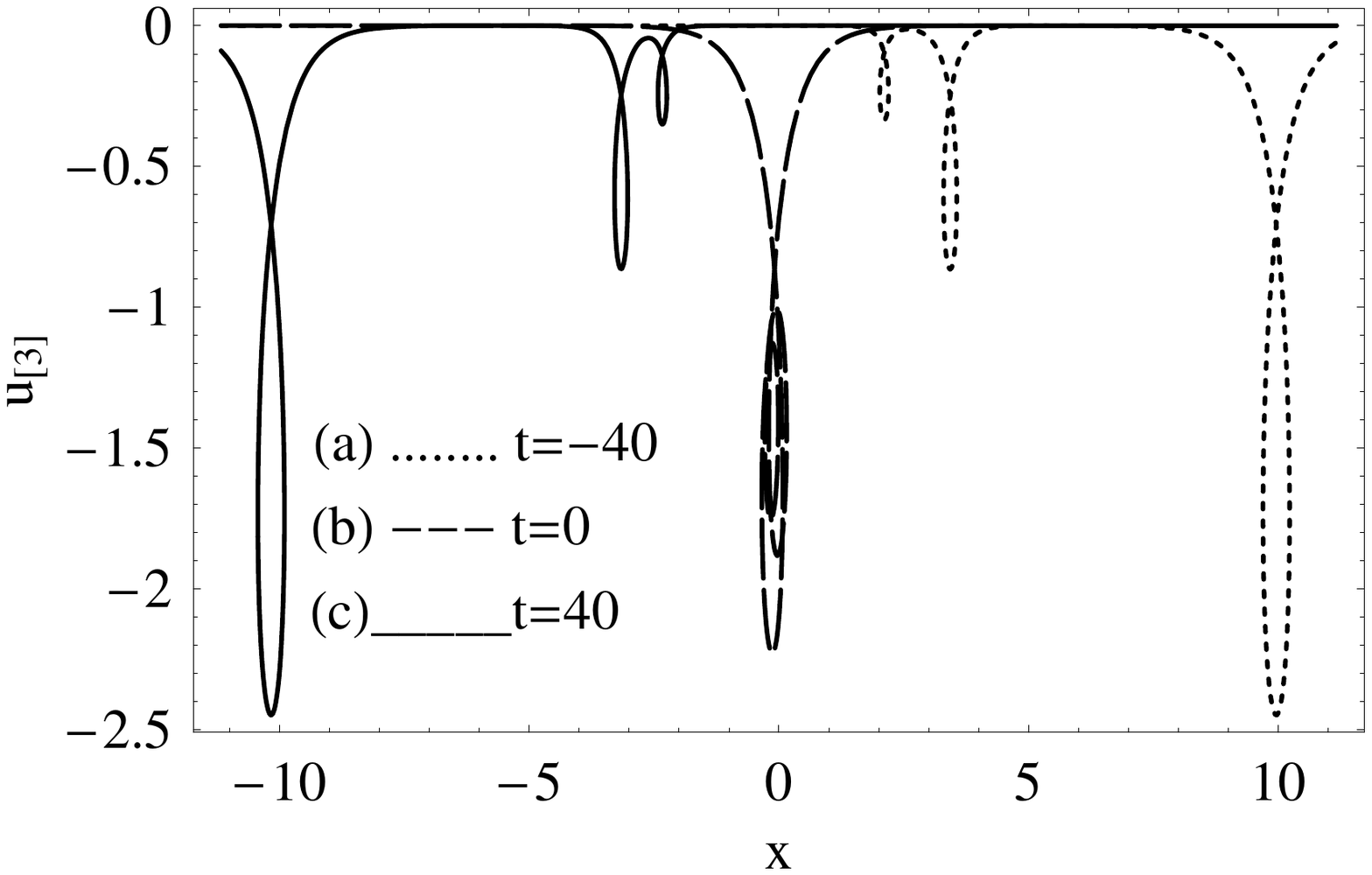}}
\resizebox{2.5in}{!}{\includegraphics{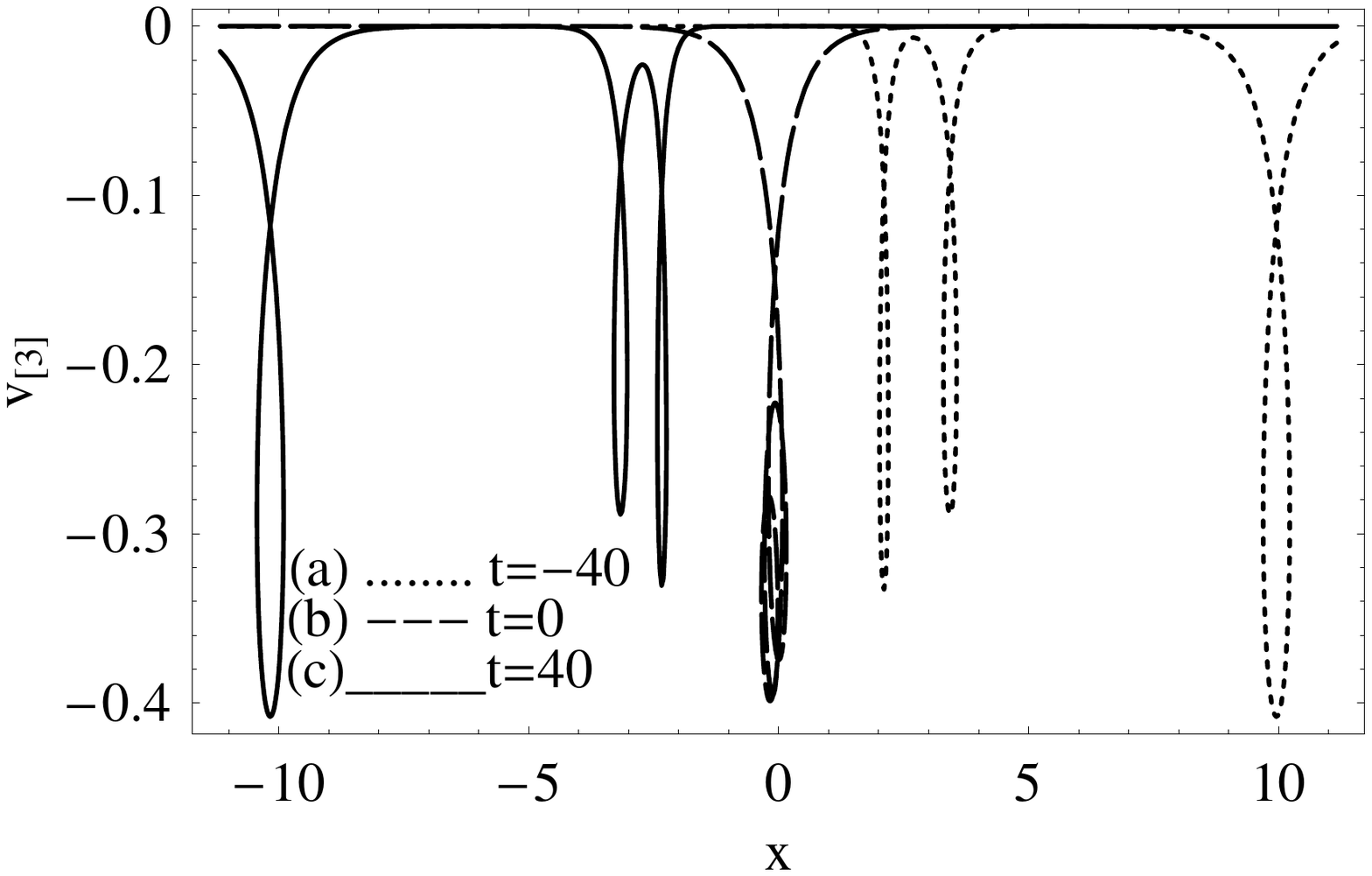}}
\end{center}
\centerline{\small{Figure 8. \,The interactional dynamics of three-soliton
solution (\ref{55}) with
$\lambda_1=-\lambda_2=3$,\,\,$\lambda_3=-\lambda_4=-2$,\,\,$\lambda_5=-\lambda_6=1$}}
\centerline{\small{$\mu_1=2,\,\,\mu_2=-\frac{1}{2}$,\,\,
$\mu_3=3$,\,\,$\mu_4=-1$,\,\,$\mu_5=4$,\,\,$\mu_6=-2$,$\alpha=1$,
 and $\beta=0$.}}
\begin{center}
\resizebox{2.5in}{!}{\includegraphics{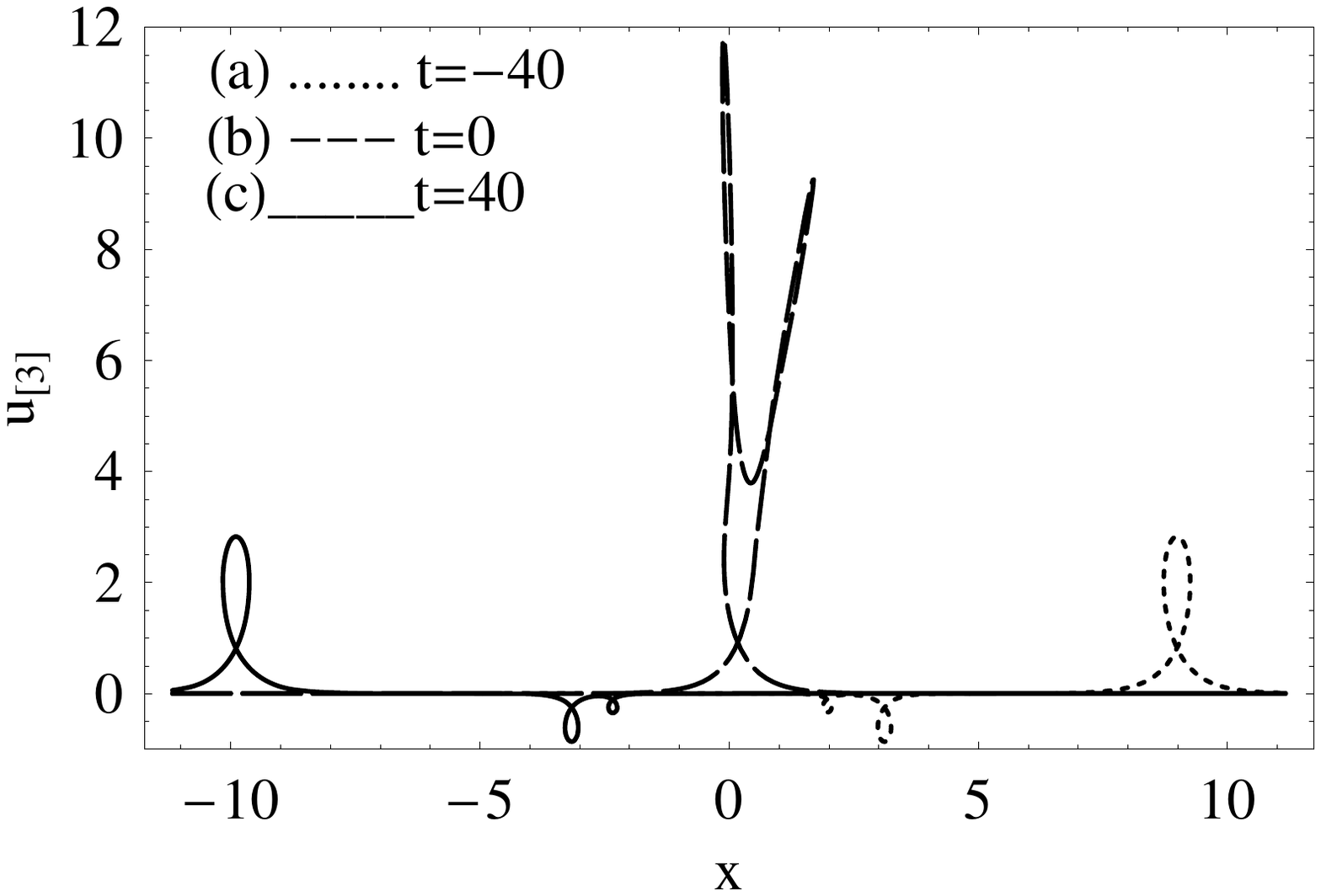}}
\resizebox{2.5in}{!}{\includegraphics{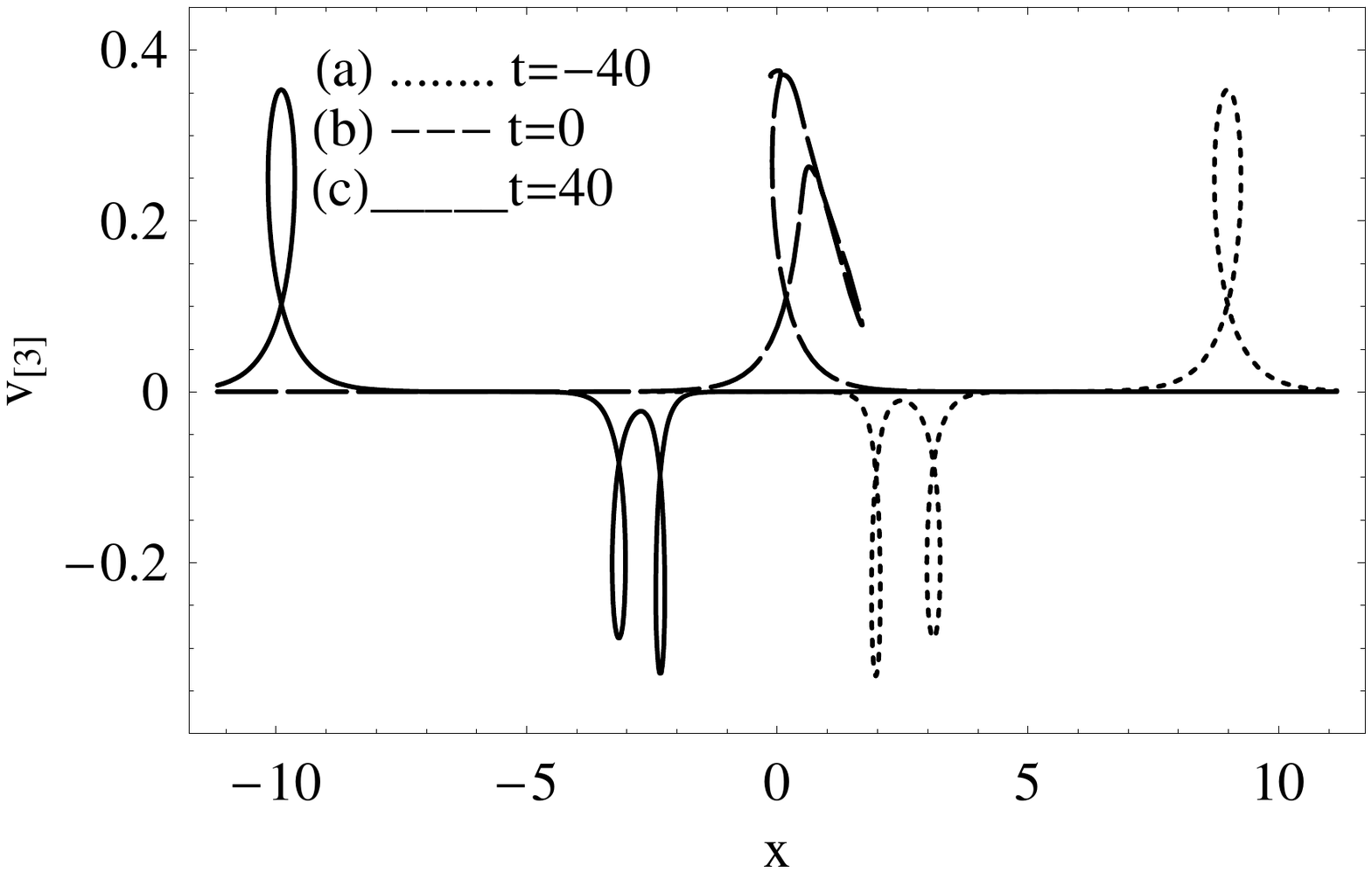}}
\end{center}
\centerline{\small{Figure 9. \,The interactional dynamics of three-soliton
solution (\ref{55}) with
$\lambda_1=-\lambda_2=3$,\,\,$\lambda_3=-\lambda_4=-2$,\,\,$\lambda_5=-\lambda_6=-1$}}
\centerline{\small{$\mu_1=2,\,\,\mu_2=-\frac{1}{2}$,\,\,
$\mu_3=3$,\,\,$\mu_4=-1$,\,\,$\mu_5=4$,\,\,$\mu_6=-2$,$\alpha=1$,
 and $\beta=x_0=0$.}}



\section{Cauchy problem for the 2SP system (\ref{2})}

Since the 2SP system (\ref{2}) is integrable, from the view point of soliton solutions, we already derived its multi-loop soliton solutions through the Darboux transformation described in the above sections. In this section, from the view point of analysis, we want to study the Cauchy problem of
the 2SP system (\ref{2}). To do so, let us present some preliminary works.
First, let us recall the definition of a scale of Banach spaces $\{X_{\delta}\}_{0<\delta\leq 1}$.
\begin{defn}
A scale of complex Banach spaces is a one-parameter family of
complex Banach spaces $\{X_{\delta}\}_{0<\delta\leq 1}$ such that:

(Scale): If for any $0 <\delta'< \delta \leq  1$ we have
\begin{equation}\label{scale}
X_{\delta}\subset X_{\delta'},~~\|\cdot\|_{\delta'}\leq \|\cdot\|_{\delta}.
\end{equation}
\end{defn}

Then, we present the framework in an analytic space. 
In the following contexts, we denote  the Fourier transform of $f$ in $\mathbb{R}$ or $\mathbb{T}$ by $\widehat{f}$. Assume that
the initial data belong to the decreasing Banach space in the following scale. For $\delta > 0$ and $s \geq 0$, in the periodic case the Banach space is defined by
\begin{equation}\label{GT}
G^{\delta,s}(\mathbb{T})=\{f\in L^2(\mathbb{T}): \|f\|_{G^{\delta,s}(\mathbb{T})}^2:=\Sigma_{k\in \mathbb{Z}}(1+|k|^{2})^{s}e^{2\delta|k|}|\hat{f}(k)|^2<\infty\}.
\end{equation}

While in the non-periodic case the Banach space is defined by
\begin{equation}\label{GR}
G^{\delta,s}(\mathbb{R})=\{f\in L^2(\mathbb{R}): \|f\|_{G^{\delta,s}(\mathbb{R})}^2:=\int_{\mathbb{R}}(1+|\xi|^{2})^{s}e^{2\delta|\xi|}|\hat{f}(\xi)|^2d\xi<\infty\}.
\end{equation}

\begin{rem}\label{rem1}
For $f\in  G^{\delta,s}$, the following properties are obvious by the definition of $G^{\delta,s}$:

(i) $0 <\delta'< \delta $ and $s \geq 0$, then $\|\cdot\|_{\delta',s}\leq \|\cdot\|_{\delta,s},$ i.e. $G^{\delta,s}\hookrightarrow G^{\delta',s}$.

(ii) $0 <s'< s $ and $\delta > 0$, then $\|\cdot\|_{\delta,s'}\leq \|\cdot\|_{\delta,s},$ i.e. $G^{\delta,s}\hookrightarrow G^{\delta,s'}$.
\end{rem}

\begin{rem}\label{rem2}
By Remark \ref{rem1} (i), it is not hard for us to see that the spaces
$$\{G^{\delta,s}\}_{0< \delta <1}, ~~with~ norm \| \cdot \|_{\delta,s}$$
form a scale of decreasing Banach spaces.
\end{rem}

Throughout the paper, 
$G^{\delta,s}$ represents the space for both the periodic and non-periodic cases if a result holds for both cases, and
$\|\cdot\|_{\delta,s}$ stands for the norm in the space $G^{\delta,s}$.





Next, let us provide some basic properties
of the $G^{\delta,s}$ spaces, which can also be seen in \cite{B-H-P1}.

\begin{lem}\label{rem3}
Let $s \geq 0$ for any $x\in\mathbb{T}$, and $s>\frac{1}{2}$ if $x\in \mathbb{R}$. Assume that $f\in  G^{\delta,s}, $ with $0<\delta\leq1$, then $f$ is a holomorphic function in a symmetric strip  $D:=\{z\in\mathbb{C}:|y|<\delta\}$, where $z=x+iy$. $\delta$ is called the analytic radius.
\end{lem}

\begin{lem}\label{rem4}
If $0 <\delta'< \delta \leq  1$, $s \geq 0$ and $f\in  G^{\delta,s}$, then
\begin{align}
\label{fx}&\|f_x\|_{\delta',s}\leq \frac{e^{-1}}{\delta-\delta'}\|f\|_{\delta,s}
\end{align}
\end{lem}

\begin{lem} (Algebraic property)\label{rem5}

If $0 <\delta< 1 $, $s>\frac{1}{2}$ and $f,g\in  G^{\delta,s}$, then we have
\begin{equation}\label{algebra}
\|fg\|_{\delta,s}\leq C_s \|f\|_{\delta,s}\|g\|_{\delta,s}.
\end{equation}
with $C_s =\sqrt{2(1+2^{2s})\sum_{k=0}^{\infty}\frac{1}{1+|k|^{2s}}} $ in the periodic case and
$C_s =\sqrt{\frac{2(1+2^{2s})}{2s-1}} $ in the non-periodic case.
\end{lem}


\begin{lem}\label{rem6}
If $u_0 \in C^w(\mathbb{T})$, there exists $\delta_0 > 0$ such that $u_0\in  G^{\delta_0,s}(\mathbb{T})$ for any $s > 0$.
\end{lem}

Furthermore, we present a brief description of the autonomous
Ovsyannikov theorem \cite{Tre1,Tre2,Tre3} that are used in the following sections.
Given a decreasing scale of Banach spaces $\{X_{\delta}\}_{0<\delta\leq 1}$ and initial data $u_0\in  X_1$, we consider the Cauchy problem
\begin{equation}\label{Fu}
\frac{du}{dt}=F(u),~~u(0)=u_0
\end{equation}
where $F:~X_{\delta}\rightarrow X_{\delta'}$ satisfies the following conditions:

(1) $F:~X_{\delta}\rightarrow X_{\delta'}$ is a function, and 
for any given $u_0 \in  X_1$ and $R>0$ there exist two positive constants $L$ and $M$, depending on $u_0$ and $R$, such that for all
$0 < \delta' < \delta < 1$ and $u_1,u_2 \in X_{\delta}$ with $\|u_1-u_0\|_{\delta}<R$ and $\|u_2-u_0\|_{\delta}<R$ we have the following Lipschitz type condition
\begin{equation}\label{con1}
\|F(u_1)-F(u_2)\|_{\delta'}\leq \frac{L}{\delta-\delta'}\|u_1-u_2\|_{\delta},
\end{equation}
and the following bound in the $X_{\delta}$ norm of $F(u_0)$
\begin{equation}\label{con2}
\|F(u_0)\|_{\delta}\leq \frac{M}{1-\delta},~0<{\delta}<1.
\end{equation}

(2) For $0<\delta'<{\delta}<1$ and $a>0$, if the function $t\mapsto u(t)$ is holomorphic on $\{t \in \mathbb{C} :
|t| < a(1 - \delta)\}$ with values in $X_{\delta}$ and $\sup_{|t|<a(1- \delta )} \|u(t) - u_0\|_{\delta} < R$, then the function
$t \mapsto F(u(t))$ is holomorphic on $\{t \in \mathbb{C} :
|t| < a(1 - \delta)\}$ with values in $X_{\delta'}$.

Let us now describe the autonomous Ovsyannikov theorem below.
\begin{thm}\cite{B-G1}\label{qq1}
(Autonomous Ovsyannikov Theorem)
 Assume that the scale of Banach spaces
$X_{\delta}$ and the function $F(u)$ satisfy the above conditions (1) and (2). Then, for given $u_0 \in X_1$ and
$R > 0$ there exists $T > 0$ such that
\begin{equation}\label{TT}
T =\frac{R}{16LR + 8M},
\end{equation}
and a unique solution $u(t)$ to the Cauchy problem (\ref{Fu}), which for every $ \delta \in (0, 1)$ is a holomorphic
function in the disc $D(0, T(1 - \delta))$ valued in $X_{\delta}$ satisfying
\begin{equation}
\sup_{|t|<T(1-\delta)} \|u(t) - u_0\|_{\delta} < R,~~ 0 < \delta < 1.
\end{equation}
\end{thm}

\subsection{Existence and uniqueness}
Let us consider the following Cauchy problem for the 2SP system (\ref{2})
\begin{equation}\label{ivpa}
\left\{\begin{array}{ll}
u_{tx}=\frac{1}{2}(uvu_x)_x+u, \\
v_{tx}=\frac{1}{2}(uvv_x)_x+v, \\
u(0,x)=u_0(x),\\v(0,x)=v_0(x).
\end{array}\right.
\end{equation}

Casting the integral operator $\partial_x^{-1}$ onto both sides of (\ref{ivpa}) yields
\begin{equation}\label{ivpa1}
\left\{\begin{array}{ll}
u_{t}=\frac{1}{2}uvu_x +\partial_x^{-1}u:=F_1(z), \\
v_{t}=\frac{1}{2}uvv_x +\partial_x^{-1}v:=F_2(z),\\
u(0,x)=u_0(x),
\\v(0,x)=v_0(x).
\end{array}\right.
\end{equation}
Putting $z=(u,v)^T, z_0=(u_0,v_0)^T$ and $ F(z)=(F_1(z),F_2(z))^T$ leads (\ref{ivpa1}) to the following concise form 
\begin{equation}
\frac{dz}{dt}=F(z),~~z(0)=z_0.
\end{equation}
Notice that there is the integral operator $\partial^{-1}_x$ involved in Eq. (\ref{ivpa1}). Let us give a remark about it below.
\begin{rem}
 $\partial_x^{-1} f$ is defined via the Fourier transform as follows
$$
\widehat{{\partial_x^{-1} f}}=\frac{1}{i\xi}\hat{f}(\xi).
$$
Due to the singularity 
at $\xi = 0$, one requires that $\hat{f}(0) = 0$,
 which is clearly equivalent to
$$
\int_{\mathbb{R}}
f(x) dx = 0.
$$
In what follows, $\partial_x^{-1} f\in L^2(\mathbb{R})$ means that there is an $ L^2(\mathbb{R})$ function $g$ such that
$g_x = f$, at least in the distributional sense.
\end{rem}


To deal with the integral term $\partial^{-1}_x u$ in Eq. (\ref{ivpa1}), we establish the following key lemma.
\begin{lem}\label{integral}
For $0 <\delta\leq 1 $, $s\geq0$ and $u\in  G^{\delta,s}$,  the following estimate holds true:
\begin{align}
\|\partial^{-1}_xu\|^2_{G^{\delta,s}}\leq 2\|u\|^2_{G^{\delta,s}}.
\end{align}
 \end{lem}
{\bf Proof: }
\begin{align*}
\|\partial^{-1}_xu\|^2_{G^{\delta,s}}&=\int_{\mathbb{R}}(1+|\xi|^{2})^{s}|e^{\delta|\xi|}\widehat{\partial^{-1}_xu}|^2d\xi\\
&=\int_{\mathbb{R}}(1+|\xi|^{2})^{s}|\frac{e^{\delta|\xi|}\widehat{u}}{i\xi}|^2d\xi\\
&=\int_{\mathbb{R}}(1+|\xi|^{2})^{s}|\widehat{\partial^{-1}_xe^{\delta(-\Delta)^{\frac{1}{2}}}u}|^2d\xi\\
&=\|\partial^{-1}_xe^{\delta(-\Delta)^{\frac{1}{2}}}u\|_{H^s}^2=\|\partial^{-1}_x f\|_{H^s}^2 \quad (Here~f=e^{\delta(-\Delta)^{\frac{1}{2}}}u )\\
&=\|\partial^{-1}_x f\|_{L^2}^2+\|\partial^{-1}_x f\|_{\dot{H}^{s}}^2 \quad (\because H^s=L^2\cap \dot{H}^s,~ \textit{with}~s\geq0)\\
&=\|f\|_{\dot{H}^{-1}}^2+\|f\|_{\dot{H}^{s-1}}^2\leq 2\|f\|_{H^s}^2=2\|e^{\delta(-\Delta)^{\frac{1}{2}}}u\|_{H^s}^2\\
&=2\int_{\mathbb{R}}(1+|\xi|^{2})^{s} |\widehat{e^{\delta(-\Delta)^{\frac{1}{2}}}u}|^2d\xi\\
&=2\int_{\mathbb{R}}(1+|\xi|^{2})^{s}e^{2\delta|\xi|}|\widehat{u}|^2d\xi=2\|u\|^2_{G^{\delta,s}}.
\end{align*}
\begin{rem}
Lemma \ref{integral} is also true for the periodic case by using the definition (\ref{GT}).
\end{rem}
For the sake of simplicity, we shall assume that
our initial data $u_0 \in G^{1,{s}}$, and for any $0<\delta<1$ and $s>\frac{1}{2}$, we define $\|z\|_{\delta,s}=\|u\|_{\delta,s}+\|v\|_{\delta,s}$.
Let us now present the existence and uniqueness of Eq. (\ref{ivpa1}).
\begin{thm}\label{qq}
Let $s > \frac{1}{2} .$ If 
$z_0=(u_0,v_0)^T \in G^{1,{s}}\times G^{1,{s}}$,
then there exists a positive
time $T$, which depends on the initial data $z_0$ and $s$, such that for every $\delta \in (0, 1),$ the Cauchy
problem (\ref{ivpa1}) has a unique solution $z(t)=(u(t),v(t))^T$. And $u(t),v(t)$ are holomorphic functions in the disc $D(0, T(1-\delta))=\{t\in\mathbb{C}:|t|<T(1- \delta)\}$ valued
in $G^{\delta,{s}}$.
 Furthermore, the analytic lifespan $T$ satisfies
 \begin{equation}\label{TTa}
T =\frac{1}{68C_s^2{e^{-1}}\|z_0\|_{1,s}^2 + 24\sqrt{2}}
\end{equation}
where $\|z_0\|_{1,s}=\|u_0\|_{1,s}+\|v_0\|_{1,s}$ and $C_s$ comes from (\ref{algebra}).
\end{thm}

Remark \ref{rem2} ensures that $G^{\delta,s}$ satisfies the scale decreasing condition (\ref{scale}) like the space $X_{\delta}$ in the autonomous
Ovsyannikov theorem. Also, these spaces and $F_i(z),i=1,2$ satisfy condition (2).
Therefore, to prove
Theorem 3.1, it suffices to show that the right-hand side $F_i(z),i=1,2$ of Eq. (\ref{ivpa1}) satisfies conditions (\ref{con1}) and (\ref{con2}). This is included in the following crucial lemma.

\begin{lem}\label{lem1} Let $s > \frac{1}{2}$. Also, let $R > 0$ and $z_0=(u_0,v_0)^T \in G^{1,s}\times G^{1,s}$
 be given. Then, for the Cauchy problem (\ref{ivpa1}) there exist positive constants $L$ and $M$, which depend on $R$ and $\|z_0\|_{1,s}$
such that for $z_1, z_ 2 \in G^{\delta,s}\times G^{\delta,s}$, $\|z_1-z_0\|_{\delta,s} < R,~ \|z_2-z_0\|_{\delta,s} < R $
and $0 < \delta'<\delta <1$ we have
 \begin{align}
~~~\|F_i(z_1)-F_i(z_2)\|_{\delta',s}
\leq \frac{L}{\delta-\delta'}\|z_1-z_2\|_{\delta,s},~~i=1,2
\end{align}
where $L=C_s^2e^{-1}(R+\|z_0\|_{1,s})^2
+\sqrt{2},$
and
\begin{align}
&~~~\|F_i(z_0)\|_{\delta,s}\leq \frac{M}
{1-\delta},~~i=1,2.
\end{align}
with
\begin{align}
M=\frac{1}{2}C_s^2e^{-1}\|z_0\|^3_{1,s}+\sqrt{2}\|z_0\|_{1,s}.
\end{align}
Moreover, the analytic lifespan $T$ satisfies the estimate
\begin{equation}
T =\frac{1}{68C_s^2{e^{-1}}\|z_0\|_{1,s}^2 + 24\sqrt{2}}.
\end{equation}
\end{lem}

{\bf Proof:}
We first prove that $F_1(z)$ satisfies (\ref{con1}) and (\ref{con2}). For $s>\frac{1}{2}$, applying the triangle inequality we get
\begin{align*}
&~~~~~\|F_1(z_1)-F_1(z_2)\|_{\delta',s}\\&\leq \frac{1}{2}\|(u_1v_1u_{1,x}-u_2v_22u_{2,x})\|_{\delta',s}
+\|\partial_x^{-1}u_{1}-\partial_x^{-1}u_{2}\|_{\delta',s}:=I+II.
\end{align*}
(\ref{fx}) and (\ref{algebra}) lead to
\begin{align}\label{I}
\nonumber I&\leq\frac{1}{2}C_s^2[\|u_1\|_{\delta',s}\|v_1\|_{\delta',s}\|(u_1-u_2)_x\|_{\delta',s}+\|u_1\|_{\delta',s}\|u_{2,x}\|_{\delta',s}\|v_1-v_2\|_{\delta',s}\\
\nonumber &~~~~+\|v_2\|_{\delta',s}\|u_{2,x}\|_{\delta',s}\|u_1-u_2\|_{\delta',s}]\\
\nonumber&\leq\frac{1}{2}C_s^2\|u_1\|_{\delta',s}\|v_1\|_{\delta',s}\frac{e^{-1}}{\delta-\delta'}\|u_1-u_2\|_{\delta,s}\\
\nonumber&~~~~+
\frac{1}{2}C_s^2\|u_{1}\|_{\delta,s}\frac{e^{-1}}{\delta-\delta'}\|u_{2}\|_{\delta,s}\|v_1-v_2\|_{\delta,s}\\
\nonumber&~~~~+ \frac{1}{2}C_s^2\|v_{2}\|_{\delta,s}\frac{e^{-1}}{\delta-\delta'}\|u_{2}\|_{\delta,s}\|u_1-u_2\|_{\delta,s}\\
&\leq\frac{1}{2}C_s^2\frac{e^{-1}}{\delta-\delta'}(R+\|z_0\|_{1,s})^2\|z_1-z_2\|_{\delta,s},
\end{align}
where we have used the following estimates
\begin{equation}\label{m1m2}
\|u_i\|_{\delta,s}\leq \|u_i-u_0\|_{\delta,s}+\|u_0\|_{\delta,s}\leq R+\|z_0\|_{1,s},~~i=1,2
\end{equation}
and
\begin{equation*}
\|v_i\|_{\delta,s}\leq \|v_i-v_0\|_{\delta,s}+\|v_0\|_{\delta,s}\leq R+\|z_0\|_{1,s},~~i=1,2.
\end{equation*}
Next, we estimate the term $II$. Lemma \ref{integral} gives
\begin{align}\label{II}
&~~~~II=|\partial_x^{-1}(u_1-u_2)\|_{\delta',s}
\leq \sqrt{2}\|(u_1-u_2)\|_{\delta',s}\leq \frac{\sqrt{2}}{\delta-\delta'}\|z_1-z_2\|_{\delta,s},
\end{align}
where we have adopted  the fact that $0<\delta'<\delta<1$, which implies $0<\delta-\delta'<1$.

Adding (\ref{I}) and (\ref{II}) gives the desired condition (\ref{con1})
\begin{align}\label{FF}
~~~\|F_1(z_1)-F_1(z_2)\|_{\delta',s}
\leq \frac{L}{\delta-\delta'}\|z_1-z_2\|_{\delta,s},
\end{align}
with $L=C_s^2e^{-1}(R+\|z_0\|_{1,s})^2
+\sqrt{2}$.

Finally, we estimate $F_1(z_0)$. (\ref{fx}) and Lemma \ref{integral} lead to
\begin{align*}
\|F_1(z_0)\|_{\delta',s}&\leq \frac{1}{2}|\|u_0v_0u_{0,x}\|_{\delta',s}+\|\partial_x^{-1}u_0\|_{\delta',s} \\
&\leq\frac{1}{2}C_s^2\|u_0\|_{\delta',s}\|v_0\|_{\delta',s}\|u_{0,x}\|_{\delta',s}+\sqrt{2}\|u_0\|_{\delta',s}\\
&\leq\frac{1}{2}C_s^2\|u_0\|_{\delta',s}\|v_0\|_{\delta',s}\frac{e^{-1}}{\delta-\delta'}\|u_0\|_{\delta,s}
+\frac{\sqrt{2}}{\delta-\delta'}\|u_0\|_{\delta,s}\\
&\leq \frac{\frac{1}{2}C_s^2e^{-1}\|z_0\|^3_{\delta,s}+\sqrt{2}\|z_0\|_{\delta,s}}{\delta-\delta'}
\end{align*}
Replacing $\delta'$ by $\delta$ and $\delta$ by 1, and setting
\begin{align*}
M=\frac{1}{2}C_s^2e^{-1}\|z_0\|^3_{1,s}+\sqrt{2}\|z_0\|_{1,s},
\end{align*}
we obtain the desired estimate (\ref{con2}), namely,
\begin{align*}
&~~~\|F_1(z_0)\|_{\delta,s}\leq \frac{M}
{1-\delta}.
\end{align*}
The symmetric structure of Eq. (\ref{ivpa1}) immediately reads 
\begin{align}\label{FF1}
~~~\|F_2(z_1)-F_2(z_2)\|_{\delta',s}
\leq \frac{L}{\delta-\delta'}\|z_1-z_2\|_{\delta,s},
\end{align}
and
\begin{align*}
&~~~\|F_2(z_0)\|_{\delta,s}\leq \frac{M}
{1-\delta},
\end{align*}
where $L$ and $M$ are the same as above. Thus, we obtain
\begin{align}
~~~\|F(z_1)-F(z_2)\|_{\delta',s}
\leq \frac{L}{\delta-\delta'}\|z_1-z_2\|_{\delta,s},
\end{align}
and
\begin{align}\label{Fz0}
&~~~\|F(z_0)\|_{\delta,s}\leq \frac{M}
{1-\delta}.
\end{align}

Therefore, substituting the above $L$ and $M$ into (\ref{TT}) yields
\begin{align*}
T &=\frac{R}{16LR + 8M}\\
&=\frac{R}{16[C_s^2e^{-1}(R+\|z_0\|_{1,s})^2+\sqrt{2}]R+8(\frac{1}{2}C_s^2e^{-1}\|z_0\|^3_{\delta,s}+\sqrt{2}\|z_0\|_{1,s})}.
\end{align*}

Thanks to Theorem \ref{qq1},
there exists a unique solution $z(t)$ to the Cauchy problem (\ref{ivpa1}), which  is
a holomorphic vector function for every $\delta \in(0, 1)$ in $D(0, T(1 - \delta)) \mapsto G^{\delta,s}\times G^{\delta,s}$ and
\begin{align*}
\sup_{|t|<T(1-\delta )} \|z(t)- z_0\|_{\delta,s} < R.
\end{align*}

Let
$R=\|z_0\|_{1,s}$, then we have
\begin{align*}
T =\frac{1}{68C_s^2{e^{-1}}\|z_0\|_{1,s}^2 + 24\sqrt{2}}.
\end{align*}
 This completes the proof of Lemma \ref{lem1}, and hence Theorem \ref{qq} is true.

\subsection{Continuity of the data-to-solution map}

\par
We now prove the continuity of the data-to-solution
map for initial data and solution in Theorem 3.1.

First, let us recall that the scale of Banach spaces $X_{\delta}$ and the function
$F(u)$ satisfy the conditions (1) and (2).  For $b > 0$, we denote by $H(|t| < b;X_{\delta}\times X_{\delta})$ the set of holomorphic vector functions $f(t)$ in $|t| < b$ valued in $X_{\delta}\times X_{\delta}$. Also, notice
that for $ 0<\delta\leq 1$ and $w \in H(|t| < b;X_{\delta}\times X_{\delta})$ with $b > 0$, the equation
\begin{equation}
\frac{dz}{dt}=F(z),~~z(0)=z_0,
\end{equation}
has a unique solution $z \in H(|t| < b;X_{\delta}\times X_{\delta})$ given by
\begin{align}
z(t)=z_0+Kw(t):=z_0+\int_0^t w(\tau)d\tau.
\end{align}
Therefore, it follows that the existence of $z$ in Theorem \ref{qq1} is equivalent to the existence of
$z \in H(|t| < T(1-\delta);X_{\delta}\times X_{\delta})$, for every $ \delta\in (0, 1)$, satisfying for $|t| < T(1-\delta)$
\begin{align}
\|\int_0^t w(\tau)d\tau\|_{\delta}<R
\end{align}
and
\begin{align}\label{w}
w=F(z_0+Kw).
\end{align}
Therefore, our initial value problem is converted to find the fixed point of the equation (\ref{w}). To see that, let us introduce a new space $E_a$.
\begin{defn}
For $a>0$ we denote by $E_a=\bigcap _{0<\delta<1} H(D(0, a(1 - \delta));X_{\delta})$ the Banach space of all functions $t\mapsto u(t)$
where for every $0<\delta<1$ we have 
\begin{align}
u: \{t:|t|<a(1 - \delta)\}\rightarrow X_{\delta}~~~~is~~holomorphic,
\end{align}
whose norm is defined by
\begin{align}\label{Ea}
|||u|||_{a} := \mathop {\sup }\limits_{\scriptstyle |t|<a(1-\delta) \hfill \atop
  \scriptstyle 0 < \delta < 1 \hfill}\{\|u(t)\|_{\delta}(1-\delta)\sqrt{1-\frac{|t|}{a(1-\delta)}} \}<\infty.
\end{align}
\end{defn}
\begin{rem}
One may easily see that $E_{T_2}\hookrightarrow E_{T_1}$ for any
$0<T_1<T_2$.
\end{rem}

\begin{rem}
 It is clear from the proof of Theorem \ref{qq} that under the hypotheses (1) and
(2) that given $z_0 \in G^{1,s}\times G^{1,s}$ and $R > 0$ there are $T > 0$ and a unique solution to the Cauchy
problem (\ref{ivpa1}) in the set
\begin{align*}
E_{T,R} := \{z(t) \in \bigcap _{0<\delta<1}
H(D(0, T(1 - \delta));G^{\delta,s}\times G^{\delta,s})\\ ~~and~ \sup_{|t|<T(1-\delta)} \|z(t) - z_0\|_{\delta,s}
< R, ~0 < \delta < 1\}.
\end{align*}
Notice that if $z\in E_{T,R}$  then $z\in E_{T}$. Thus, this allows us endow $E_{T,R}$ with the metric
$d(z, w) = |||z -w|||_T $.
\end{rem}

Using the spaces $E_a$ and the norm (\ref{Ea}) we may readily obtain the following two lemmas. We refer the readers to \cite{B-H-P1} for the detailed proofs.
\begin{lem}\label{Ku}
Let $a>0$, $u\in E_a$, $0 < \delta < 1$ and $|t|<a(1-\delta)$. Then
\begin{align*}
\|Ku(t)\|_{\delta}\leq \int_0^{|t|}\|u(\tau \frac{t}{|t|})\|_{\delta} d\tau\leq 2a|||u|||_a.
\end{align*}
\end{lem}

\begin{lem}\label{aaa}
For every $a>0$, $u\in E_a$, $0 < \delta < 1$ and $|t|<a(1-\delta)$, we have
\begin{align*}
\int_0^{|t|}\frac{\|u(\tau \frac{t}{|t|})\|_{\delta(\tau)}}{\delta(\tau)-\delta} d\tau\leq \frac{8a|||u|||_a}{1-\delta}\sqrt{\frac{a(1-\delta)}{a(1-\delta)-|t|}},
\end{align*}
where $\delta(\tau)=\frac{1}{2}(1+\delta-\frac{|\tau|}{a})$.
\end{lem}


Next, let us recall the following definition of the continuity of the data-to-solution map for Eq. (\ref{ivpa1}).
\begin{defn}\cite{B-H-P1}\label{MT}
 One says that the data-to-solution map $z_0\mapsto z(t)$ is continuous if for a given
$z_{\infty}(0) \in G^{1,s}\times G^{1,s}$ there exist $T = T({\|z_{\infty}(0)\|_{1,s}} )> 0$ and $R > 0$ such that for any sequence
of initial data $z_{n}(0) \in G^{1,s}\times G^{1,s}$ converging to $z_{\infty}(0)$ in $G^{1,s}\times G^{1,s}$ the corresponding solutions, $z_{n}(t),z_{\infty}(t)$ to the Cauchy problem (\ref{ivpa1}) for all sufficiently large $n$ satisfy: $z_{n}(t),z_{\infty}(t)\in E_{T,R}$ and $|||z_{n}(t)-z_{\infty}(t)|||_T=|||u_{n}(t)-u_{\infty}(t)|||_T+|||v_{n}(t)-v_{\infty}(t)|||_T\rightarrow 0$, where
\begin{align*}
|||u|||_T:=\sup\{\|u(t)\|_{\delta}(1-\delta)\sqrt{1-\frac{|t|}{T(1-\delta)}}:0<
\delta<1,|t|<T(1-\delta)\}<\infty.
\end{align*}
\end{defn}

We now give the continuity of the solution map for the Cauchy problem (\ref{ivpa1}).
\begin{thm}\label{cc}
Given $z_0 \in G^{1,s}\times G^{1,s}, s > \frac{1}{2}$, and $R > 0$ there exists $T = T_{z_0} > 0$, which is
given in (\ref{TT}), such that the Cauchy problem for (\ref{ivpa1}) has a unique solution
\begin{align*}
z\in E_{T,R} :=\{ z(t) \in \bigcap _{0<\delta<1}
H(D(0, T(1 - \delta));G^{\delta,s}\times G^{\delta,s}) \\~~and~ \sup_{|t|<T(1-\delta)} \|z(t) - z_0\|_{\delta}
< R, ~0 < \delta < 1\}.
\end{align*}
Moreover the data-to-solution map $z_0\mapsto z(t): G^{1,s}\times G^{1,s}\mapsto E_{T,R}$ is continuous.
\end{thm}

{\bf Proof }
Let $s > \frac{1}{2}$, $z_{\infty}(0)\in G^{1,s}\times G^{1,s}$ be given. And let $z_{n}(0)\in G^{1,s}\times G^{1,s}$ be a sequence of initial data converging to $z_{\infty}(0)$, that is $\|z_{n}(0)-z_{\infty}(0)\|_{1,s}\rightarrow 0$, as $n\rightarrow \infty$. Therefore, there exists a natural integer  $N\in\mathbb{N}$, such that for any $n\geq N$, we have
\begin{align}\label{un}
\|z_{n}(0)\|_{1,s}\leq\|z_{\infty}(0)\|_{1,s}+1.
\end{align}
Setting
\begin{align}\label{RR}
R_{\infty}=\|z_{\infty}(0)\|_{1,s}+1,
 \end{align}
 and for $n>N$
 \begin{align}\label{RRR}
\nonumber R_{n}&=R_{\infty}+\|z_{n}(0)-z_{\infty}(0)\|_{1,s}\\
&\leq R_{\infty}+1.
 \end{align}
 For the given initial data $z_{\infty}(0),z_{n}(0)\in G^{1,s}\times G^{1,s}$, Theorem \ref{qq} ensures the existence and uniqueness of the corresponding solutions $z_{\infty}(t)\in E_{T_{z_{\infty}(0)},R_{\infty}}$ and $z_{n}(t)\in E_{T_{z_{n}(0)},R_{n}}$ with
\begin{align}\label{m1}
z_{\infty}(t)=z_{\infty}(0)+K(F(z_{\infty}(t))),~~for~|t|<T_{z_{\infty}(0)}(1-\delta),
\end{align}
\begin{align}\label{m2}
z_{n}(t)=z_{n}(0)+K(F(z_{n}(t))),~~for~|t|<T_{z_{n}(0)}(1-\delta),
\end{align}
respectively,
where their lifespans are given by
  \begin{align*}
&~~~~T_{z_{\infty}(0)} =\frac{1}{68C_s^2e^{-1}\|z_{\infty}(0)\|_{1,s}^2+24\sqrt{2}},
\end{align*}
  \begin{align*}
&~~~~T_{z_{n}(0)} =\frac{1}{68C_s^2e^{-1}\|z_{n}(0)\|_{1,s}^2+24\sqrt{2}},
\end{align*} respectively.

Let us now figure out the same lifespan of $z_{\infty}(t)$ and $z_{n}(t)$ by setting $T$ as follows
  \begin{align*}
&~~~~T_{z_{\infty}(0),z_{n}(0)} =\frac{1}{68C_s^2e^{-1}R^2+24\sqrt{2}},
\end{align*}
where
\begin{align}\label{RRRR}
R=2R_{\infty}+1.
\end{align}
(\ref{un}), (\ref{RR}) and (\ref{RRRR}) imply that $T_{z_{\infty}(0),z_{n}(0)}< T_{z_{\infty}(0)}$ and $T_{z_{\infty}(0),z_{n}(0)}< T_{z_{n}(0)}$, that is $T_{z_{\infty}(0),z_{n}(0)}< \min\{ T_{z_{\infty}(0)}, T_{z_{n}(0)}\}. $

For $n\geq N$, noticing $E_{T_{z_{\infty}(0), R_{\infty}}}\hookrightarrow E_{T_{z_{\infty}(0),z_{n}(0)},R}$ and $E_{T_{z_{n}(0), R_n}}\hookrightarrow E_{T_{z_{\infty}(0),z_{n}(0)},R}$ gives 
$z_{\infty}(t),z_{n}(t)\in E_{T_{z_{\infty}(0),z_{n}(0)},R}$.

Next, we need to prove $|||z_n - z_{\infty}|||_{T_{z_{\infty}(0),z_{n}(0)}}\rightarrow 0$ as $n\rightarrow \infty$.
For $0<\delta<1$ and $|t|<T_{z_{\infty}(0),z_{n}(0)}(1-\delta)$ it follows from (\ref{m1}) and (\ref{m2}) that
\begin{align}
||z_n - z_{\infty}||_{\delta,s}  \leq \|K[F(z_n(t))-F(z_{\infty}(t))]\|_{\delta,s}+\|z_n(0) - z_{\infty}(0)\|_{\delta,s} .
\end{align}

Using Lemma \ref{Ku} (the complete proof of the lemma can be found in Lemma 6 \cite{B-H-P1}), we have
\begin{align*}
\int_0^t \|F(z_{\infty}(y))-F(z_{n}(y))\|_{\delta,s}dy\leq
\int_0^{|t|}\|F(z_{\infty}(\tau \frac{t}{|t|}))-F(z_{n}(\tau \frac{t}{|t|}))\|_{\delta,s} d\tau,
\end{align*}
and therefore, we get
\begin{align}\label{mm}
\nonumber&\|z_n(t) - z_{\infty}(t)\|_{\delta,s}-\|z_n(0) - z_{\infty}(0)\|_{\delta,s}\\
&\leq
\int_0^{|t|}\|F(z_{\infty}(\tau \frac{t}{|t|}))-F(z_{n}(\tau \frac{t}{|t|}))\|_{\delta,s} d\tau.
\end{align}

In order to use (\ref{con1}), we also need to prove that
\begin{align*}
\|z_{\infty}(\tau \frac{t}{|t|}) - z_{n}(0)\|_{\delta(\tau),s}<R,
\end{align*}
and
\begin{align*}
\|z_{n}(\tau \frac{t}{|t|}) - z_{n}(0)\|_{\delta(\tau),s}<R.
\end{align*}
Apparently, if $0<\delta<1$ and $|t|<T_{z_{\infty}(0),z_{n}(0)}(1-\delta)$, $0\leq |\tau|=\tau \leq |t|$, $\delta<\delta(\tau)\leq\frac{1}{2}(1+\delta-\frac{|\tau|}{T_{z_{\infty}(0),z_{n}(0)}})$
and $n>N$, then we have
\begin{align}\label{mn2}
\nonumber&~~~~\|z_{\infty}(\tau \frac{t}{|t|}) - z_{n}(0)\|_{\delta(\tau),s}\\
\nonumber &\leq \|z_{\infty}(\tau \frac{t}{|t|}) - z_{\infty}(0)\|_{\delta(\tau),s}+\|z_{\infty}(0) - z_{n}(0)\|_{\delta(\tau),s}\\
&<R_{\infty}+1<R.
\end{align}
Theorem \ref{qq1} and (\ref{RRR}) ensure that
\begin{align}\label{mn22}
\|z_{n}(\tau \frac{t}{|t|}) - z_{n}(0)\|_{\delta(\tau),s}\leq R_{n}<R.
\end{align}
(\ref{mn2}), (\ref{mn22}), (\ref{con1}) and Lemma \ref{aaa} lead to
\begin{align}\label{mm1}
\nonumber&\|z_n(t) - z_{\infty}(t)\|_{\delta,s}-\|z_n(0) - z_{\infty}(0)\|_{\delta,s}\\
\nonumber&\leq
\int_0^{|t|}\|F(z_{\infty}(\tau \frac{t}{|t|}))-F(z_{n}(\tau \frac{t}{|t|}))\|_{\delta,s} d\tau\\
&\leq
L_n
\int_0^{|t|}\frac{\|z_{\infty}(\tau \frac{t}{|t|})-z_{n}(\tau \frac{t}{|t|}))\|_{\delta(\tau),s}}{\delta(\tau)-\delta} d\tau
\end{align}
with $L_n=C_s^2e^{-1}(R_n+\|z_0\|_{1,s})^2
+\sqrt{2}$ and $\delta(\tau)=\frac{1}{2}(1+\delta-\frac{|\tau|}{T_{z_{\infty}(0),z_{n}(0)}})$. Noticing for
$|\tau|<T_{z_{\infty}(0),z_{n}(0)}(1-\delta)$, we have $0<\delta<\delta(\tau)<1$.

As per Lemma \ref{aaa} (with $a=T_{z_{\infty}(0),z_{n}(0)}$), and (\ref{mm1}), for $0<\delta<1$ and $|t|<T_{z_{\infty}(0),z_{n}(0)}(1-\delta)$ we obtain
\begin{align*}
&\|z_n(t) - z_{\infty}(t)\|_{\delta,s}-\|z_n(0) - z_{\infty}(0)\|_{\delta,s}\\
&\leq
\frac{8T_{z_{\infty}(0),z_{n}(0)}L_n|||z_{\infty}-z_n|||_{T_{z_{\infty}(0),z_{n}(0)}}}
{1-\delta}\sqrt{\frac{T_{z_{\infty}(0),u_{n}(0)}(1-\delta)}{T_{z_{\infty}(0),z_{n}(0)}(1-\delta)-|t|}},
\end{align*}
which reveals that
\begin{align*}
\nonumber &~~~|||z_n(t) - z_{\infty}(t)|||_{T_{z_{\infty}(0),z_{n}(0)}}\\
&\leq8T_{z_{\infty}(0),z_{n}(0)}L_n|||z_{\infty}-z_n|||_{T_{z_{\infty}(0),z_{n}(0)}}+\|z_n(0) - z_{\infty}(0)\|_{\delta,s}
\end{align*}
in turns implies that
\begin{align}
\nonumber&~~~(1-8T_{z_{\infty}(0),z_{n}(0)}L_n)|||z_n(t) - z_{\infty}(t)|||_{T_{z_{\infty}(0),z_{n}(0)}}\\
&\leq\|z_n(0) - z_{\infty}(0)\|_{\delta,s}
\end{align}
For $n>N$, (\ref{RR}), (\ref{RRR}) and (\ref{RRRR}) give
\begin{align}\label{ln}
\nonumber L_n&=C_s^2e^{-1}(R_n+\|z_n(0)\|_{1,s})^2
+\sqrt{2}\\
\nonumber&\leq C_s^2e^{-1}(R_n+\|z_{n}(0)-z_{\infty}(0)\|_{1,s}+\|z_{\infty}(0)\|_{1,s})^2
+\sqrt{2}\\
\nonumber&\leq C_s^2e^{-1}(R_n+1+\|z_{\infty}(0)\|_{1,s})^2
+\sqrt{2}\\
\nonumber&= C_s^2e^{-1}(R_n+R_{\infty})^2
+\sqrt{2}\\
&\leq C_s^2e^{-1}R^2
+\sqrt{2}
\end{align}
Noticing $T_{z_{\infty}(0),z_{n}(0)}=\frac{1}{68C_s^2e^{-1}R^2+24\sqrt{2}}$ and (\ref{ln}), we have \begin{align*}
8T_{z_{\infty}(0),z_{n}(0)}L_n\leq\frac{8C_s^2e^{-1}R^2
+8\sqrt{2}}{68C_s^2e^{-1}R^2+24\sqrt{2}}
<\frac{1}{2},
\end{align*} which implies that
 \begin{align}
 \nonumber &~~~|||z_n(t) - z_{\infty}(t)|||_{T_{z_{\infty}(0),z_{n}(0)}}\\
 \nonumber&\leq\frac{1}{1-8T_{z_{\infty}(0),z_{n}(0)}L_n}\|z_{\infty}-z_n\|_{1,s}\\
&\leq 2\|z_n(0) - z_{\infty}(0)\|_{1,s}.
  \end{align}
This completes the proof of Theorem \ref{cc}.

\begin{rem}
Theorem \ref{cc} showed  important results since it makes the 2SP system (\ref{ivpa}) to be well-posed
in the spaces $G^{\delta,s}\times G^{\delta,s}$
in the sense of Hadamard. One may compare these results with the
classical Cauchy-Kovalevski theorem, where there is no continuity of the data-to-solution map.
\end{rem}



\section{Conclusions}

In this paper, we have used the Darboux transformation to
solve the 2SP system (\ref{2}) with loop, anti-loop, and multi-loop soliton solutions.
Those solutions are explicitly given and graphically depicted through the plotted graphs.
The approach utilized in this paper may be developed in the following ways:

(i) In light of the transformation (\ref{5}) and Darboux
transformation (\ref{413}), we may obtain the following  parametric
representation of soliton solutions to equations (\ref{6})
\begin{equation}\label{60}
\begin{array}{ll}
x[N]=x[0]+\dfrac{\Delta_{A_{N-1}}}{\Delta_{N-1}},\\[3mm]
q=\dfrac{1}{2}\left(u[0]+v[0]+\dfrac{\Delta_{B_{N-1}}+\Delta_{C_{N-1}}}{\Delta_{N-1}}\right),\\[3mm]
r=-\dfrac{{\rm
i}}{2}\left(u[0]-v[0]+\dfrac{\Delta_{B_{N-1}}-\Delta_{C_{N-1}}}{\Delta_{N-1}}\right),
\end{array}
\end{equation}
where $\Delta_{N-1}$, $\Delta_{A_{N-1}}$, $\Delta_{B_{N-1}}$, and
$\Delta_{C_{N-1}}$ are given through (\ref{424})-(\ref{427}).

(ii) The $N$-soliton solution of the
short pulse equation (\ref{1}) could be regained as a reduction of the Darboux transformation (\ref{413}) with (\ref{44}) under the conditions
$\lambda_{2k}=-\lambda_{2k-1}$ and $\mu_{2k}\,\mu_{2k-1}=-1$
$(k=1,2,\cdots,N)$.

(iii) Our 2SP system (\ref{2}) could be extended to a two-component complex short pulse model, which we shall pursue a deeper study in the near future.

In the analysis aspect of the  2SP system (\ref{2}),
we adopted the abstract Ovsyannikov type theorem and proved the well-posedness of the Cauchy problem for the 2SP system (\ref{2}) provided that the initial data are analytic.
It is very interesting to study the existence and the uniqueness of the global weak solutions, which we will discuss elsewhere.  Furthermore, as mentioned above, the analysis work for the complex short pulse model is also deserved to investigate.

\section*{Acknowledgments}
This work is partially supported by the National Natural Science Foundation of
China under (Grant Numbers: 11261037, 11401223, and 11171295), the Natural Science Foundation of
Inner Mongolia Autonomous Region under (Grant No 2014MS0111), the Natural Science Foundation of Guangdong (No. 2015A030313424), the
Caoyuan Yingcai Program of Inner Mongolia Autonomous Region under
(Grant No CYYC2011050), the Program for Young Talents of Science and
Technology in Universities of Inner Mongolia Autonomous Region under
(Grant No NJYT14A04). This work is also partially aided by an overseas funding grant from China
Scholarship council and the Haitian Scholar Plan of Dalian University of Technology.


\end {document}